\documentclass[a4paper,11pt]{article}
\pdfoutput=1 

\usepackage{jheppub} 

\usepackage{graphicx,epsfig,wrapfig,amssymb,color,amsmath,bm,breqn}

\usepackage[T1]{fontenc} 

\allowdisplaybreaks

\usepackage{lineno} 

\usepackage{amssymb}
\usepackage{amsmath}
\usepackage{graphicx}
\usepackage{float}
\usepackage{caption}
\usepackage{soul}
\usepackage{slashed} 
\usepackage{color}
\usepackage{mathtools}
\usepackage{bm}
\usepackage{enumerate}
\usepackage{subcaption}

\newcommand{\no}{\nonumber \\}

\newcommand{\hadpsc}{\ensuremath{P_{\rm B}}}

\newcommand{\hadmass}{\ensuremath{M_{\rm B}}}
\newcommand{\hadtmass}{ M_{\rm B, T}}
\newcommand{\pmass}{\ensuremath{M}}

\newcommand{\parz}[1]{\ensuremath{\left(#1\right)}}

\newcommand{\smallm}{\ensuremath{m}}

\newcommand{\order}[1]{\ensuremath{O\parz{#1}}}
\newcommand{\contractor}[1]{\ensuremath{{\rm P}}_{#1}}
\newcommand{\contractortot}[1]{\ensuremath{{\rm P}}_{#1}}

\newcommand{\contractortotap}[1]{\ensuremath{{\rm \tilde{P}}}_{#1}}

\newcommand{\Tsc}[2]{#1_{#2\rm{\scriptsize T}}}
\newcommand{\Tscsq}[2]{#1^2_{#2\rm{T}}}
\newcommand{\T}[2]{\ensuremath{{\bf #1}}_{#2\text{\scriptsize T}}}

\newcommand\3[1]{\boldsymbol{#1}}

\newcommand{\xbj}{\ensuremath{x_{\rm Bj}}}
\newcommand{\xn}{\ensuremath{x_{\rm N}}}
\newcommand{\xh}{\ensuremath{x_{\rm h}}}
\newcommand{\zh}{\ensuremath{z_{\rm h}}}
\newcommand{\zex}{\ensuremath{z_{\rm N}}}

\newcommand{\xnp}{\ensuremath{\hat{x}_{\rm N}}}

\newcommand{\zexp}{\ensuremath{\hat{z}_{\rm N}}}
\newcommand{\eref}[1]{Eq.~(\ref{e.#1})}
\newcommand{\erefs}[2]{Eqs.~(\ref{e.#1})--(\ref{e.#2})}
\newcommand{\fref}[1]{Fig.~\ref{f.#1}}

\newcommand{\aref}[1]{Appendix~\ref{a.#1}}
\newcommand{\sref}[1]{Sec.~\ref{s.#1}}

\newcommand{\mom}[3]{\ensuremath{{#1}^{#3}_{#2}}}

\newcommand{\momt}[3]{\ensuremath{{{\bf #1}^{{#3}}_{#2 \rm{T}}}}} 
\newcommand{\momtsc}[3]{\ensuremath{{{\bf #1}^{{#3}}_{#2 \rm{T}}}}} 

\newcommand{\hadmom}[2]{\ensuremath{\mom{P}{{\rm B}, #2}{#1} }}
\newcommand{\hadmomap}[2]{\ensuremath{\mom{\tilde{P}}{{\rm B}, #2}{#1} }}
\newcommand{\hadmomt}[1]{\ensuremath{\momt{P}{{\rm B}, #1, }{}}}
\newcommand{\hadmomtap}[1]{\ensuremath{\momt{\tilde{P}}{\rm B, #1, }{}}}
\newcommand{\hadmomplain}{\ensuremath{P_{\rm B}}}
\newcommand{\hadmomplaint}{\ensuremath{\momt{P}{{\rm B},}{}}}
\newcommand{\hadmomtsc}[2]{\ensuremath{\momtsc{P}{{\rm B}, #2}{#1} }}
\newcommand{\hadmomtsq}[2]{\ensuremath{\momt{P}{{\rm B}, #2,}{#1} }}

\newcommand{\qmom}[2]{  \ensuremath{\mom{q}{#2}{#1} }}
\newcommand{\qmomt}[1]{\ensuremath{\momt{q}{#1}{}}}
\newcommand{\qmomtsc}[2]{\ensuremath{\momtsc{q}{#2}{#1} }}
\newcommand{\Pmom}[2]{\ensuremath{\mom{P}{#2}{#1} }}
\newcommand{\Pmomt}[1]{\ensuremath{\momt{P}{#1}{}}}

\newcommand{\genmomt}[2]{\ensuremath{\momt{#1}{\rm b, #2, }{}}}
\newcommand{\initq}{\ensuremath{k_{\rm i}}}
\newcommand{\initqb}{\ensuremath{k_{\rm i, b}}}
\newcommand{\xmom}{\ensuremath{k_{\rm X}}}
\newcommand{\finalq}{\ensuremath{k_{\rm f}}}
\newcommand{\finalqb}{\ensuremath{k_{\rm f, b}}}
\newcommand{\initqtb}{\ensuremath{{\bf k}_{\rm i, b, T}}}
\newcommand{\initqt}{\ensuremath{{\bf k}_{\rm i, T}}}

\newcommand{\finalqt}{\ensuremath{{\bf k}_{\rm f, T}}}

\newcommand{\finalqtb}{\ensuremath{{\bf k}_{\rm f, b, T}}}
\newcommand{\finalqtH}{\ensuremath{{\bf k}_{\rm f, H, T}}}

\newcommand{\py}{\ensuremath{y_{P, {\rm b}}}}
\newcommand{\hady}{\ensuremath{y_{\rm B, b}}}
\newcommand{\inity}{\ensuremath{y_{\rm i}^{\rm b}}}
\newcommand{\finaly}{\ensuremath{y_{\rm f}^{\rm b}}}
\newcommand{\ratcur}{\ensuremath{R_1}}
\newcommand{\rattar}{\ensuremath{R'_1}}
\newcommand{\kinep}{\ensuremath{\varepsilon}}
\newcommand{\ratgen}{\ensuremath{R_0}}
\newcommand{\ratTM}{\ensuremath{R_2}}
\newcommand{\ratX}{\ensuremath{R_3}}

\DeclareRobustCommand*\diff[2][]{%
   \mathop{
     \mathrm{d}^{#1}
     \mskip-0.2\thinmuskip
    #2}\nolimits
}

\preprint{JLAB-THY-19-2920}

\title{	Mapping the  Kinematical Regimes of Semi-Inclusive Deep Inelastic Scattering}

\author[a]{M.~Boglione}
\emailAdd{elena.boglione@to.infn.it}
\affiliation[a]{Dipartimento di Fisica, Universit\`a di Torino, INFN-Sezione Torino,
  Via P. Giuria 1, 10125 Torino, Italy}
  
\author[b]{A.~Dotson}
\emailAdd{adots004@nmsu.edu}
\affiliation[b]{Department of Physics, New Mexico State University, Las Cruces, NM, 88003, USA}  

\author[c]{L.~Gamberg}
\emailAdd{lpg10@psu.edu}
\affiliation[c]{Science Division, Penn State University Berks, Reading, 
Pennsylvania 19610, USA}

\author[d]{S.~Gordon}
\emailAdd{ssgordon@odu.edu}
\affiliation[d]{Department of Physics, Old Dominion University, Norfolk, VA 23529, USA}
\affiliation[e]{Jefferson Lab, 12000 Jefferson Avenue, Newport News, VA 23606, USA}

\author[a]{J.~O.~Gonzalez-Hernandez}
\emailAdd{joseosvaldo.gonzalez@to.infn.it}
 
\author[e,c]{A.~Prokudin}
\emailAdd{prokudin@jlab.org}

\author[e,d]{T.~C.~Rogers}
\emailAdd{trogers@jlab.org}
\author[e]{N.~Sato}
\emailAdd{nsato@jlab.org}

\abstract{
We construct a language for identifying kinematical regions of
transversely differential semi-inclusive deep inelastic scattering cross sections
with particular underlying partonic pictures, especially in regions of
moderate to low $Q$ where sensitivity to kinematical effects outside
the usual very high energy limit becomes non-trivial. The partonic
pictures map to power law expansions whose leading contributions ultimately lead to well-known QCD factorization theorems. We propose methods for estimating the consistency 
of any particular region of overall hadronic kinematics with the kinematics of 
a given underlying partonic picture.
The basic setup of kinematics of semi-inclusive deep inelastic
scattering is also reviewed in some detail.}

\keywords{
	semi-inclusive deep inelastic scattering, quantum chromo dynamics, kinematics, transverse momentum dependent distribution and fragmentation
	functions}

\begin{document}


\maketitle

\flushbottom
\newpage

\section{Introduction}
\label{s.intro}

Deep inelastic scattering (DIS), and especially semi-inclusive deep
inelastic scattering (SIDIS) are headlining processes in most programs
to the study of partonic (quark and gluon) degrees of freedom. It is a
cornerstone process of, for example, the Jefferson Lab 12 GeV  program to
study partonic structure in hadrons, and is one of the important
processes for study in a possible future Electron-Ion Collider (EIC)
~\cite{Dudek:2012vr,Accardi:2012qut,Aschenauer:2014twa,Avakian:2014hga,Aschenauer:2016our,Avakian:2016rst}.
Interest in SIDIS arises from a variety of considerations.
Well-established collinear factorization theorems for SIDIS provide
access to the flavor dependence of standard parton distribution functions 
(PDFs) and fragmentation functions (FFs).  In the target fragmentation function region,
different kinds of objects, called fracture 
functions~\cite{Trentadue:1993ka,Graudenz:1994dq}, are involved and these are 
sensitive to still other novel QCD phenomena.  Beyond collinear factorization, 
transversely differential SIDIS at low transverse momentum is sensitive to the 
properties of transverse momentum dependent (TMD) PDFs and FFs. 

Many DIS experiments are performed at medium $Q$ (roughly
$1$-$3$~GeV), where it is reasonable to expect some sensitivity to
intrinsic properties of hadron structure and other 
non-perturbative effects to be significant. 
By ``moderate-to-low $Q$,'' we will mean SIDIS measurements with $Q$ roughly 
between $1$~GeV and $3$~GeV and Bjorken-$\xbj$ not too far below the 
valence region. This includes JLab 6 GeV and 12 GeV SIDIS cross section 
measurements~\cite{Dudek:2012vr,Avakian:2014hga,Avakian:2016rst}.  

So, the moderate-to-low $Q$ region has some obvious   
advantages in the mission to refine the current view of 
hadron structure. 
If all energies and hard scales are extremely large, then
asymptotic freedom means that 
pictures of partonic interactions rooted in perturbation theory 
can usually be 
applied confidently and with very high accuracy and precision.  But, with the large relative 
fraction of the hard process contributions and perturbatively produced radiation involved, 
it becomes less clear to 
what extent observables are truly sensitive to the intrinsic
properties of the actual hadron constituents. This further
points to moderate-to-low $Q$ measurements as ideal sources 
of information about partonic hadron structure.
However, there are also unique challenges to interpreting 
moderate-to-low $Q$ cross sections, particularly for less inclusive versions 
of DIS like SIDIS. With lower hard scales, access to intrinsic
effects of constituents may be more direct, 
but this also comes with less confidence in the reliability and accuracy
of perturbative and/or parton-based descriptions. 
Moreover, the average final state hadron multiplicity in such measurements is
typically less than about 3 in the valence region of Bjorken-$\xbj$.
In long term efforts to establish
intrinsic properties for partons, the trade-off in advantages at large and small $Q$ 
needs to be 
confronted systematically, and such that knowledge of one complements  
the other. 

Sophisticated theoretical frameworks, usually involving some form of
QCD factorization and perturbation 
theory~\cite{Muta:1987mz,Greiner:2002ui,Collins:2011qcdbook} have long existed for describing 
specific underlying physical mechanisms in many highly differential processes over many regions, including in SIDIS,
in terms of partonic degrees
of freedom.  However, they always assume specific kinematical limiting cases, e.g. very large or very small transverse momentum, or very large
or very small rapidity.  The interface between different physical
regimes remains somewhat unclear in practice, and most especially when
the hard scales involved are not especially large.  Estimating the
kinematical boundaries of any specific QCD approach or approximation beyond very rough orders of magnitude is difficult and subtle. It requires at least some model assumptions, e.g. about
the role of parton virtuality and/or the onset of various non-perturbative or
hadronic mechanisms generally. Monte Carlo simulations can help, but these 
also involve physical assumptions whose impact needs to be understood 
systematically. Future phenomenological and experimental efforts will hopefully 
clarify the location of region boundaries, and discriminate between competing 
hypotheses.

The aim of the present article is to discuss how such questions can be
posed in a systematic way.  To this end, we will refrain from
discussing specific theoretical frameworks or QCD models and instead
enumerate the steps needed to map any given set of assumptions
concerning exact intrinsic partonic/constituent properties to a corresponding
kinematical region of $\xbj$, $Q$, $\zh$, and transverse momentum in a
cross section. The goal is to organize an interpretation strategy applicable with any 
model of underlying non-perturbative dynamics for
exact parton momentum, independent even of assumptions about factorization.
Our final result is a sequence of tests that probe the 
proximity of any given kinematical configuration to a conventional partonic region of SIDIS, and 
that also probe the sensitivity to the various model assumptions needed to make such an assessment. A convenient web interface 
for implementing these tests can be found at Ref.~\cite{sidistool}. 

We also review the basics of the SIDIS process itself, in some cases
translating past results into an updated language, motivated by
current research efforts.   For other general introductions to SIDIS in pQCD see, for example, Refs.~\cite{Meng:1991da,Levelt:1993ac,Meng:1995yn,Mulders:1996dh,Nadolsky:1999kb,muldersnotes}. See
especially Refs.~\cite{Anselmino:2005nn,Bacchetta:2006tn,Bacchetta:2008xw,Anselmino:2011ch} for another review
of the basics of SIDIS that includes a full catalogue of spin and
azimuthal dependences. 
For general treatments of SIDIS in the context of fracture functions and
target fragmentation, see Ref.~\cite{Graudenz:1994dq}.
Finally, see Chapters 12-13 of Ref.~\cite{Collins:2011qcdbook},
which influences much of the language and notation of this article.

The first half of this paper reviews many of the basics of SIDIS: \sref{lcvar} and 
\sref{process} provide an overview of our notation and setup, \sref{refs} discusses 
the various reference frames commonly used, \sref{fsm} explains the kinematical
characterization of final state hadron momentum, and \sref{xsecs} explains 
our conventions for the decomposition of cross sections into structure functions. 
We then begin the discussion of standard approximations in the second half of the 
paper, starting with the purely kinematical approximations in \sref{kins}. We explain the 
characterization of partonic kinematics, and the typical approximations associated 
with them, in \sref{partons}, with a focus on the current and large transverse 
momentum regions. In \sref{rapidity} we translate these 
considerations into the language of rapidity. In \sref{target} and \sref{soft} we discuss 
the target and soft regions. We provide examples of the region characterization in 
\sref{examples}, with experimentally reasonable kinematics. Finally, we make 
concluding remarks in \sref{conclusion}. 

\newpage

\section{Light-Cone Variables}
\label{s.lcvar}

Light-cone variables are defined as follows: for a four-vector $V^\mu$,
\begin{equation}
V^{\mu} = \parz{V^+, V^-, \T{V}{}} \, , 
\end{equation}
where 
\begin{equation}
V^+ = \frac{V^0 + V^z}{\sqrt{2}}, \qquad V^- 
	= \frac{V^0 - V^z}{\sqrt{2}}, \qquad  \T{V}{} = \parz{V^x, V^y}\, .
\end{equation}
For a four-momentum $V$, rapidity is defined as usual:
\begin{equation}
y = \frac{1}{2} \ln \parz{\left| \frac{V^+}{V^-} \right| } \, .
\end{equation}
In terms of rapidity, light-cone momentum is:
\begin{equation}
V = \parz{\frac{\Tsc{M}{}}{\sqrt{2}} e^y, \frac{\Tsc{M}{}}{\sqrt{2}} e^{-y}, \T{V}{} } \, ,
\end{equation}
where $V^2 = M^2$ and transverse mass is
\begin{equation}
\Tsc{M}{}  = \sqrt{\left| M^2 + \T{V}{}^2 \right| } \, .
\end{equation}
For a virtual momentum, $M^2 < 0$ and either the plus or minus light-cone component
is negative, e.g., 
\begin{equation}
V = \parz{\frac{\Tsc{M}{}}{\sqrt{2}} e^y, - \frac{\Tsc{M}{}}{\sqrt{2}} e^{-y}, \T{V}{} } \, .
\end{equation}
In labeling a four-momentum component of $V$, we will write:
\begin{equation}
\mom{V}{\rm b,c}{\rm a}{} \, , \label{e.notation}
\end{equation}
where $a$ is the contravariant component, $c$ specifies the reference
frame, and $b$ is any other necessary subscript depending on the given
context. A two-dimensional transverse momentum is
\begin{equation}
\genmomt{V}{\rm c}{} \, .
\end{equation}
The frame subscripts $b,c$ on a four-momentum indicate in which frame its
components will be expressed.
\section{The Process}
\label{s.process}

We consider the process:
\begin{equation}
\text{lepton} (l) + \text{proton} (P) 
  \to 
\text{lepton} (l') + \text{Hadron} (\hadmomplain{}{}) + X \, .
\label{e.theproc}
\end{equation}
The final state hadron has type $B$. The ``$X$" is an instruction to
sum over all unobserved particles including other $B$ hadrons. Note
that each $B$ in an event is counted. A sketch is shown in
\fref{diagram}\footnote{In this figure we followed the so-called Trento conventions~\cite{Bacchetta:2004jz}.}.

The proton has momentum $P$, the virtual photon has momentum $q$, the
produced hadron has momentum $\hadpsc$, and the incoming and scattered
leptons have momenta $l$ and $l^\prime$ respectively. The mass of the
target hadron is $\pmass$ and the mass of the produced hadron is
$\hadmass$.  
\begin{figure}[t]
\includegraphics[scale=.5]{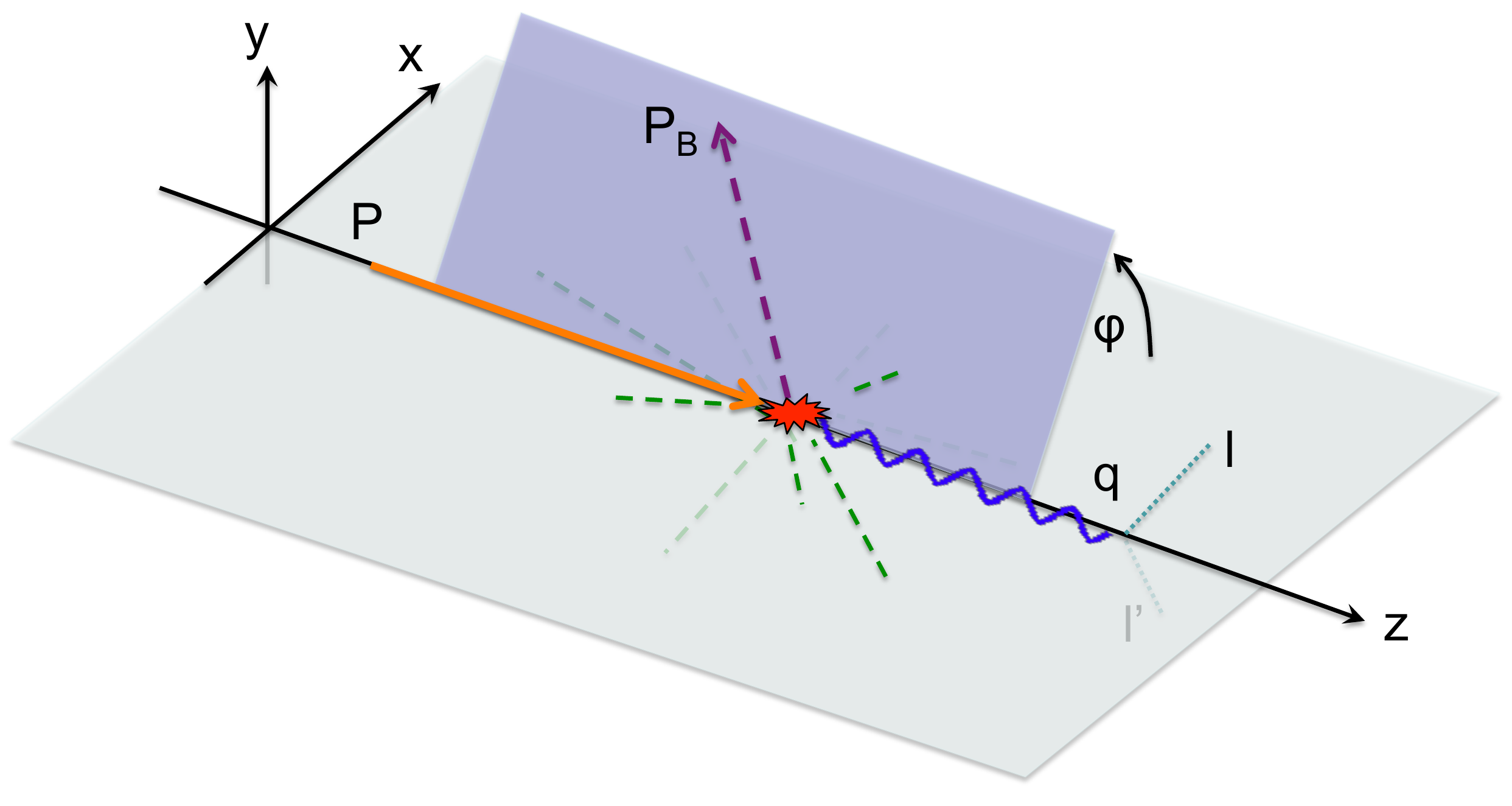}
\caption{Diagram of a SIDIS event in a photon frame - see \sref{refs}. The hadron plane 
is shown in purple. The dashed green lines represent unobserved particles.}
\label{f.diagram}
\end{figure}

Observables like cross sections or structure functions are
conventionally parameterized by a combination of the following
kinematical variables:
\begin{align}
 Q^2 &= -q^2 = -(l - l^\prime)^2\,  , \quad \quad
	\xbj = \frac{Q^2}{2 P \cdot q}\,   , \quad\quad
	  \xn  = -\frac{q^+}{P^+} =\frac{2 \xbj}{1 + \sqrt{1 +\frac{4 \xbj^2 \pmass^2}{Q^2}}}\, ,  
		\label{e.xndef} \\
  y &= \frac{P \cdot q}{P \cdot l}\,  ,  \quad\quad\quad\quad
		   \zh = \frac{P \cdot \hadpsc}{P \cdot q} = 2 \xbj \frac{P \cdot \hadpsc}{Q^2}\, ,  \quad\quad\quad\quad
                   \zex = \frac{P_{\rm B}^-}{q^-}\,   , 
	              \label{e.zhdef} \\ \nonumber\\ 
W_{\rm tot}^2  &=  \parz{q + P}^2,  \quad\quad\quad
                W_{\rm SIDIS}^2  =  \parz{q + P - \hadpsc}^2  ,  \quad\quad\quad
                       s =  (l + P)^2
                       \label{e.Wxhdef}\, .  
\end{align}
In the light-cone ratios that define $\xn$ and $\zex$, momentum
components $q^\pm$, $P^+$ and $P_{\rm B}^-$ are defined in a photon
frame (see Section~\ref{s.refs}), where the incoming proton is in the positive $z$-direction with zero transverse momentum and the virtual photon is the the negative
$z$-direction with no transverse momentum. (See  the discussion of
photon frames below.)  Since boosts along the $z$-axis do not affect
light-cone ratios, the exact photon frame does not matter. $\xn$ is
the kinematical variable usually called Nachtmann-$x$. It is often
labeled by a $\xi$, but for us $\xi$ will label a partonic momentum
fraction, so we use $\xn$ instead, with the subscripts on $\xn$ and
$\xbj$ distinguishing between Bjorken and Nachtmann $x$-variables.  
For descriptions of fragmentation, the light-cone fraction $\zex$ is
the analogue of $\xn$, and the ${\rm N}$ subscript is meant to
emphasize this analogy.  Our $\xbj$, $\zh$ and $P_{\rm B}$
correspond, respectively, to $x$, $z$ and $P_h$
from~\cite{Bacchetta:2006tn}.  Our $\zex$ corresponds to $\zeta_h$ of
Ref.~\cite{Guerrero:2015wha}.  A variable 
\begin{equation}
\xh = \frac{q \cdot \hadpsc}{P \cdot q} \label{e.xh}
\end{equation}
is useful if the target fragmentation region is being described.

The deep inelastic limit is $\smallm/Q \to \infty$ with fixed $\xn$
and $\zex$.  The ``$m$" symbol will always represent a generic mass
scale in this paper, considered to be very small relative to $Q$, such as a small
hadron mass or $\Lambda_{\rm QCD}$.  The kinematical variables obey
\begin{align}
Q^2 &{}= \xbj y (s - M^2 - m_l^2) \approx \xbj ys \, .
\end{align}
The "$\approx$" symbol will always mean "dropping $m/Q$
power-suppressed corrections" with $\xn$ and $\zex$ fixed.

\section{Reference Frames}
\label{s.refs}

It is useful and common to 
switch between photon and hadron frames. Here we describe the SIDIS kinematics 
in these frames.

\begin{itemize}
\item \underline{Photon frame}:
\end{itemize}
In a photon frame, the virtual photon and the initial proton both have
zero transverse momentum, while the final state produced hadron has
non-zero transverse momentum:
\begin{align}
\qmom{}{\gamma} & = \parz{-\xn \Pmom{+}{\gamma}, 
\frac{Q^2}{2 \xn \Pmom{+}{\gamma}} , \T{0}{} } \, ,  \label{e.qgamma} \\
\Pmom{}{\gamma} & = \parz{\Pmom{+}{\gamma}, 
\frac{M^2}{2 \Pmom{+}{\gamma}}, \T{0}{}} \, , \\ 
\hadmom{}{\gamma} & 
= \parz{ \frac{\hadmomtsc{2}{\gamma,  }  
+ \hadmass^2}{2 \hadmom{-}{\gamma}}, \hadmom{-}{\gamma}, \hadmomt{\gamma} } \, .
\end{align}
The $\gamma$ subscript signals the use of components in the photon
frame, following the notation of \eref{notation}.  In the photon frame
\begin{equation}
\label{eq:zapprox}
\hadmom{-}{\gamma} = \frac{\zh Q^2}{4 \xbj \Pmom{+}{\gamma} } 
\parz{1 \pm \sqrt{1 - \frac{4 \xbj^2 \pmass^2 \parz{ \hadmomtsc{2}{\gamma, } 
+ \hadmass^2} }{\zh^2 Q^4 } } }
\approx \frac{\zh Q^2}{2 \xbj \Pmom{+}{\gamma}} \, ,
\end{equation} 
where the approximation symbol shows the limit of zero hadron masses
for the solution corresponding to the current fragmentation region.
Note that \eref{qgamma} fixes $\xn$ to be defined as in \eref{xndef}.

The angles $\psi$ and $\phi$ are the azimuthal angles of the final
state lepton and produced hadron respectively in a photon frame. Note
that ratios of plus and minus components are independent of boosts in
the $z$-direction, and so are the same in all photon frames.

\pagebreak

\begin{itemize}
\item \underline{Breit frame}:
\end{itemize}
A particular case of the photon frame is the Breit (Brick Wall) frame, see \fref{frames}(a), where 
\begin{align}
\qmom{}{\rm b} &{}= \parz{-\frac{Q}{\sqrt{2}}, \frac{Q}{\sqrt{2}}, \T{0}{} } \, ,  
\label{e.qbreit} \\
\Pmom{}{\rm b} &{}= \parz{\frac{Q}{\xn \sqrt{2}}, \frac{\xn \pmass^2 }{\sqrt{2} Q}, \T{0}{} } 
= \parz{\frac{\pmass}{\sqrt{2}} \; e^{\py},\frac{\pmass}{\sqrt{2}} e^{-\py},\T{0}{}} \, .  
\label{e.pmombreit}
\end{align}
The small ${\rm b}$ indicates that components
are in the Breit frame. This will be our default frame, so any
four-momentum components without a subscript should be assumed to be
in the Breit frame.
\begin{figure}
\centering
  \begin{tabular}{c@{\hspace*{5mm}}c}
    \includegraphics[scale=0.5]{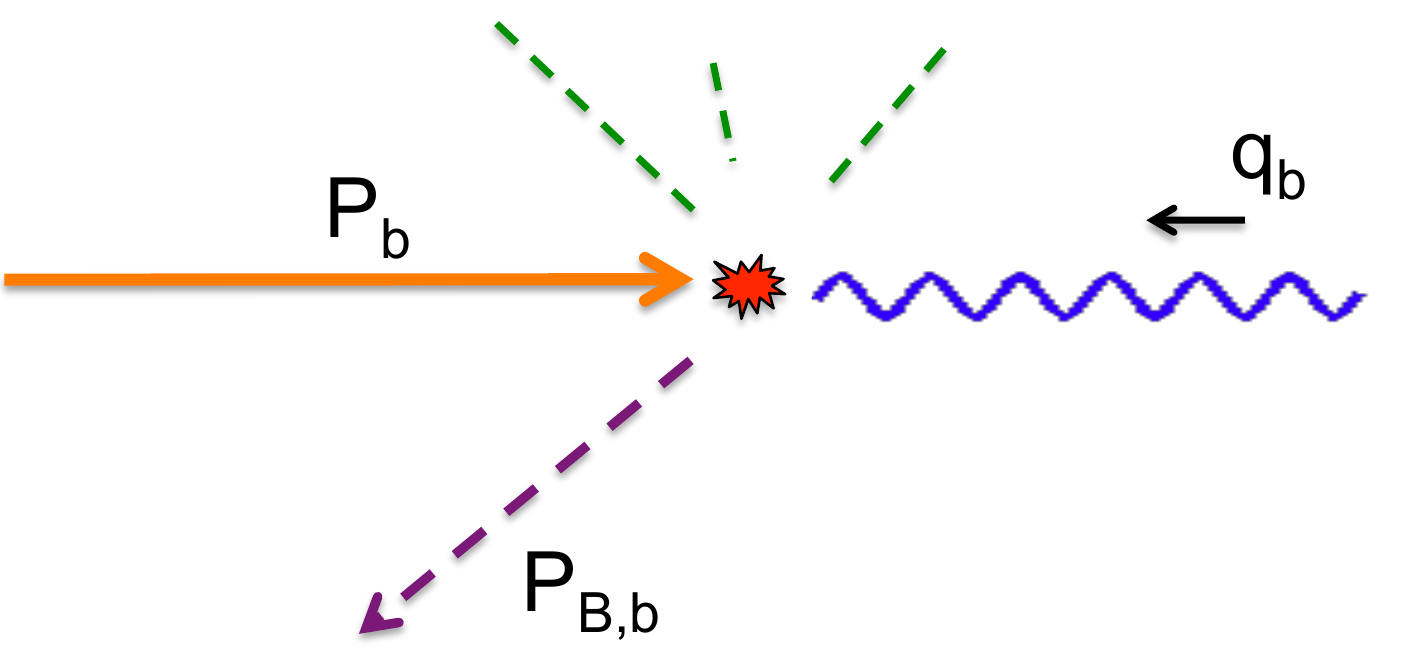}
    \hspace{0.5cm}
    &
    \hspace{0.5cm}
    \includegraphics[scale=0.5]{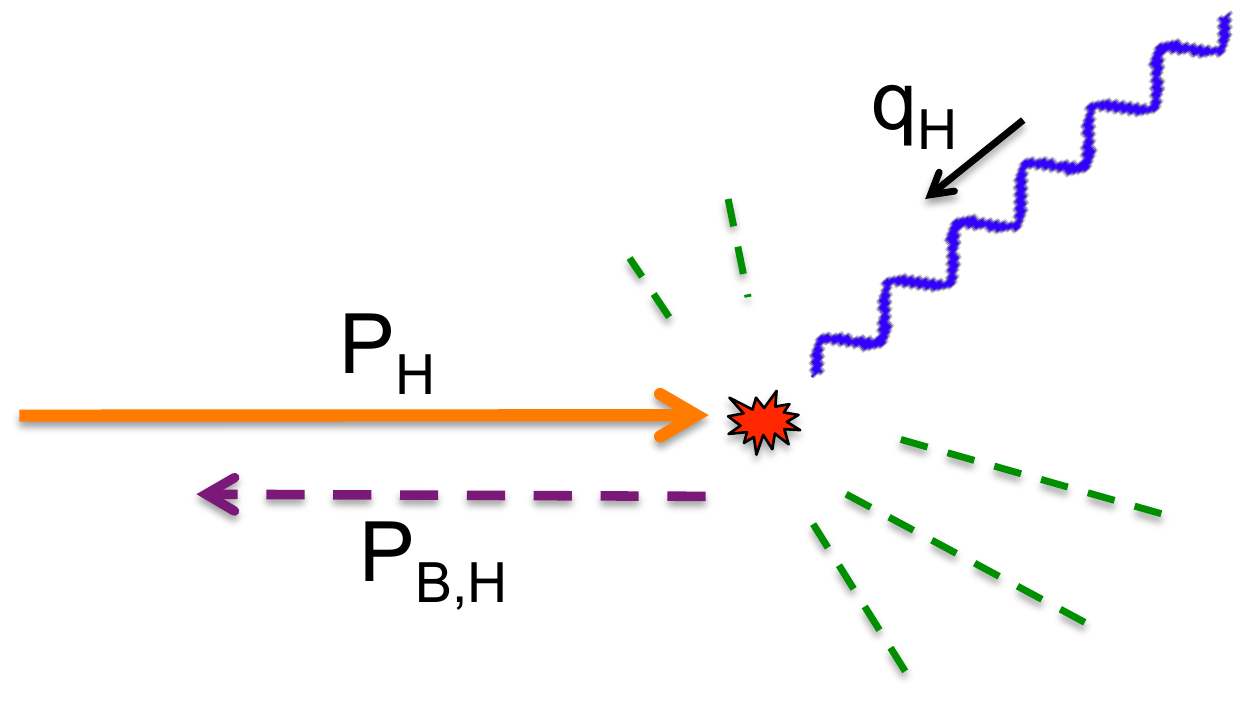}
  \\
  (a) & (b)
  \end{tabular}
\caption{
  The configuration of the proton, photon, and outgoing hadron in (a) the
  Breit photon frame and (b) the hadron frame. The dashed green lines
  again represent unobserved particles.
}
\label{f.frames}
\end{figure}


\begin{itemize}
\item \underline{Hadron Frame}:
\end{itemize}
In the hadron frame, see \fref{frames}(b), labeled by ``$\rm H$,'' the incoming hadron and
final state hadron are exactly back-to-back (zero relative transverse
momentum) while the virtual photon generally has non-zero transverse
momentum. It is an especially useful frame for setting up
factorization. (See~\cite[Sec.13.15.5]{Collins:2011qcdbook}.) The components of
the four-momenta are:
\begin{align}
\qmom{}{\rm H} & = \parz{ \qmom{+}{\rm H}, \qmom{-}{\rm H}, \qmomt{\rm H,} } \, , \\
\Pmom{}{\rm H} & = \parz{\Pmom{+}{\rm H}, \frac{M^2}{2 \Pmom{+}{\rm H}}, \T{0}{}} \, , \\ 
\hadmom{}{\rm H} & = \parz{\frac{\hadmass^2} {2 \hadmom{-}{\rm H}},
	\hadmom{-}{\rm H}, \T{0}{} } \, .
\end{align}
For definiteness, define the hadron frame such that the components of
the incoming target momentum are exactly the same as in the Breit
frame:
\begin{align}
\Pmom{+}{\rm H} &{}= \Pmom{+}{\gamma} = \frac{Q}{\sqrt{2} \xn}\, ,\\
\Pmom{-}{\rm H} &{}= \Pmom{-}{\gamma} = \frac{\xn \pmass^2}{\sqrt{2} Q}\, ,\\
\Pmomt{\rm H,} &{}= \T{0}{} \, . 
\end{align}

\begin{itemize}
\item \underline{Varying Conventions}:
\end{itemize}
For us, the hadron frame has zero transverse momentum for the
produced hadron and non-zero transverse momentum for the virtual
photon~\cite{Mulders:1996dh,Collins:2011qcdbook}. Note that this is \emph{opposite} the situation 
in the hadron frame of Meng-Olness-Soper (MOS) Ref.~\cite{Meng:1991da}. The 
MOS hadron frame corresponds to the photon frame of Collins~\cite{Collins:2011qcdbook}. MOS
define a Lorentz invariant four-vector (\cite[Eq.~(10)]{Meng:1991da})
that measures the deviation from the back-to-back configuration.
From~\cite[Eq.~(11)]{Meng:1991da} and~\cite[Eq.~(13.104)]{Collins:2011qcdbook},
the Ref.~\cite{Collins:2011qcdbook} hadron frame $\qmomtsc{2}{\rm H,}$ is the same
as the MOS $\qmomtsc{2}{\rm H,}$ if hadron masses are neglected.

Restricting to the MOS hadron frame, MOS use
\cite[Eq.~(11)]{Meng:1991da} and \cite[Eq.~(13)]{Meng:1991da} and
$\hadmomplain^2 = 0$ to find \cite[Eq.~(12)]{Meng:1991da}, which in
light-cone coordinates is \eref{hadronpara} with $\hadmass^2 = 0$ and
with the MOS $\T{q}{}$ defined to point along the positive $x$-axis.
In the MOS hadron frame, the transverse part of $\hadpsc$ is always in
the $x$ direction and is always positive.

Mulders and Tangerman~\cite[Eqs.~(15-17)]{Mulders:1996dh} give general
expressions for four vector components that include the effects of
hadron masses,  and the reference frames used correspond to the 
hadron and/or photon frames defined above. References such
as~\cite{Bacchetta:2004jz,Diehl:2005pc,Bacchetta:2006tn} specialize
the photon frame to the target rest frame rather than the Breit frame.
$\hadmomt{\rm b}$ is invariant, however, with respect to boosts along
the $z$-axis.  Other conventions use some combination of the above.
Refs.~\cite{Ji:2004xq,Idilbi:2004vb,Ji:2004wu} use a hadronic tensor
with an extra $1/4\zh$ relative to the above
and Refs.~\cite{Mulders:1996dh,muldersnotes} have an extra $1/2\pmass$.  The
notation of Ref.~\cite{Koike:2006fn} is similar to Ref.~\cite{Meng:1991da}.

\section{Variables for the final state momentum}
\label{s.fsm}
Here we treat the final state momentum in terms of the the light-cone momentum 
fraction $\zex$~\eref{zhdef} variable, and also express it in both the photon and 
hadron frames;  in turn we relate it to the hadron frame photon transverse 
momentum.

First, define the exact Breit frame $\hadmom{}{\rm b}$ in terms of $\zex$
\begin{equation}
\hadmom{}{\rm b} 
= \parz{\frac{\hadmass^2 
+ \zex^2 \qmomtsc{2}{} }{\sqrt{2} \zex Q}, \frac{\zex Q}{\sqrt{2}}  , - \zex \qmomt{} } 
=  \parz{\frac{\hadtmass}{\sqrt{2}} \; e^{\hady},\frac{\hadtmass}{\sqrt{2}} \; e^{-\hady}, \hadmomt{\rm b} } \, ,
\label{e.hadronpara}
\end{equation}
where $\qmomt{}$ is so far only a symbol used to define the Breit
frame transverse component, and it has yet to be related to physical
quantities.  In other words, first define 
\begin{equation}
\zex \equiv \frac{\hadmom{-}{\rm b}}{\qmom{-}{\rm b}} \, , \label{e.zratdef}
\end{equation}
in accordance with \eref{zhdef}, and then define
\begin{equation}
\qmomt{} \equiv -\frac{\hadmomt{\rm b}}{\zex} = -\frac{\qmom{-}{\rm b} \hadmomt{\rm b}}{\hadmom{-}{\rm b}} \, . \label{e.qtratdef}
\end{equation}
Note the minus sign. The momentum fraction $\zex$ is related to the
kinematical parameter $\zh$ by
\begin{equation}
\zex = \frac{ Q^4 \xn \zh \left( 1 
\pm \sqrt{1 - \frac{4 \pmass^2 \hadmass^2 \xbj^2 (Q^4  
+ \xn^2 \pmass^2 \qmomtsc{2}{} )}{Q^8 \zh^2} } \right)}{2 \xbj (Q^4 
+ \xn^2 \pmass^2 \qmomtsc{2}{})} 
\stackrel{\rm{Fixed} \, \xn, \zh, \Tsc{q}{}{}}{=} 
\zh \parz{1 + \order{\frac{m^4}{Q^4}} } \, . \label{e.zrels}
\end{equation}
The expansion after the far right equals sign is for the "$+$"
solution, which in conventional treatments of SIDIS corresponds to the
current fragmentation region. Note that the relationship between $\zex$
and $\zh$ is generally double valued. Parameterizing final hadron momentum as
in \eref{hadronpara} is convenient for some purposes such as in
factorization derivations. 

It is often useful to switch back and forth between the photon (e.g.,
Breit) and hadron frames.  For this, define
\begin{equation}
\kappa \equiv \sqrt{\zex^2 \qmomtsc{2}{} + \frac{\pmass^2 \xn^2 \parz{\hadmass^2 
+ \qmomtsc{2}{} \zex^2 - \frac{Q^4 \zex^2 }{\pmass^2 \xn^2} }^2 }{4 Q^4 \zex^2} } 
= \order{\frac{Q^2}{m}} \, .
\end{equation}
The law to transform a vector $V$ from the Breit frame to the hadron frame is then\footnote{The simplest sequence of transformations to get this are:  1) boost from the Breit frame to the proton rest frame 2) 
  rotate until the momentum of the final state hadron is along the negative $z$-axis 3) boost along the $z$-axis to a frame where the proton has a light-cone plus component equal to that of Breit frame $P^+_{\rm b}=Q/\xn \sqrt{2}$.}
\begin{eqnarray}
V_{\rm H}^+ &=& \frac{1}{2 \pmass^2 \xn^2} \parz{\pmass^2 \xn^2 \parz{1 
+ \sqrt{1 - \frac{\zex^2 \qmomtsc{2}{}}{\kappa^2}} } V_{{\rm b}}^+ 
- Q^2 \parz{-1 + \sqrt{1 - \frac{\zex^2 \qmomtsc{2}{}}{\kappa^2} } } V_{{\rm b}}^- } \nonumber \\
& &+ \frac{Q \zex}{\sqrt{2} \pmass \xn \kappa} \qmomt{} \cdot \momt{V}{{\rm b},}{} \, , \label{e.Vtrans} \\
V_{\rm H}^- &=& -\frac{1}{2 Q^2} \parz{\pmass^2 \xn^2 \parz{-1 
+ \sqrt{1 - \frac{\zex^2 \qmomtsc{2}{}}{\kappa^2}} } V_{{\rm b}}^+ 
- Q^2 \parz{1 + \sqrt{1 - \frac{\zex^2 \qmomtsc{2}{}}{\kappa^2} } } V_{{\rm b}}^- }\nonumber \\ 
 & &- \frac{M \xn \zex}{\sqrt{2} Q \kappa} \qmomt{} \cdot \momt{V}{{\rm b},}{} \, , \\
\momt{V}{\rm H,}{} &=& \momt{V}{{\rm b},}{} 
\sqrt{1 - \frac{\zex^2 \qmomtsc{2}{}}{\kappa^2}} 
   + \qmomt{} \frac{\zex \parz{Q^2 V_{{\rm b}}^-  
   - \pmass^2 \xn^2 V_{{\rm b}}^+ } }{\sqrt{2} \pmass Q \xn \kappa} \, . 
\end{eqnarray}
In the limit that masses are small relative to $Q$, these become
\begin{align}
V_{\rm H}^+ &{}\approx  V_{{\rm b}}^+ + \frac{\qmomtsc{2}{}}{Q^2} V_{{\rm b}}^- 
+ \frac{\sqrt{2}}{Q} \qmomt{} \cdot \momt{V}{{\rm b},}{}   \, , \\
V_{\rm H}^- &{}\approx V_{{\rm b}}^-   \, , \\
\momt{V}{\rm H,}{} &{}\approx \momt{V}{{\rm b},}{}
   + \qmomt{} \frac{\sqrt{2} V_{{\rm b}}^- }{Q} \, . 
   \label{e.lortransapp}
\end{align}
Also, using \eref{qbreit} in \eref{lortransapp},
\begin{equation}
\qmomt{\rm H,} \approx \qmomt{} \, .
\end{equation}
Comparing this with \eref{hadronpara} and using \eref{zrels} confirms that
\begin{equation}
\qmomt{\rm H,} \approx - \frac{\hadmomt{\rm b}}{\zh}   
\approx \qmomt{} \,.  \label{e.qtapps}
\end{equation}
As usual, the $\approx$ symbol means "neglecting
$m/Q$-suppressed corrections." Thus, $\qmomt{}{}$ has the physical
meaning in the limit of $m/Q \to 0$ of the hadron frame photon
transverse momentum $\qmomt{\rm H,}{}$, which in turn is
$-\hadmomt{\rm b}/\zh$. 

It is also useful to write the hadron momentum directly in terms of
its Breit frame transverse momentum:
\begin{equation}
\hadmom{}{\rm b} = \parz{\frac{M_{\rm B, T}^2 }{2 \zex \qmom{-}{\rm b}}, \zex \qmom{-}{\rm b}  , \hadmomt{\rm b}{} } \, .
\label{e.hadronpara2}
\end{equation}
The momentum fraction $\zh$ is related to the kinematical parameter
$\zex$ by
\begin{equation}
\zh = \frac{\xbj \zex}{\xn} \left(1 + \frac{\xn^2 \pmass^2 M_{\rm B, T}^2 }{\zex^2 Q^4}  \right) 
\, .
\end{equation}
The inverse is
\begin{align}
\zex &{}= \frac{\xn \zh}{2 \xbj} \left(1 + \sqrt{1 - \frac{4 \pmass^2 M_{\rm B, T}^2 \xbj^2}{Q^4 \zh^2}  } \right) \approx \zh \, . \label{e.zrels2}
\end{align}
See also \cite[Eq.~(2.12)]{Guerrero:2015wha}.  Note that $\xn$ is a
function of $\xbj$, $Q$, and $\pmass$, but we will sometimes keep it to minimize the size of expressions as
in \eref{zrels2} rather than
writing everything explicitly in terms of $\xbj$, $Q$, and $\pmass$. 

\section{Cross Sections and Structure Functions}
\label{s.xsecs}
      
A cross section differential in $N$ final state particles for particle
$A$ scattering from particle $B$ is related to modulus-squared matrix
elements $|M|^2$ in the usual way:
\begin{align}
\diff{\sigma}{} &{}= \frac{|M^{A,B \to N}|^2}{2 \lambda(s,m_A^2,m_B^2)^{1/2} } 
\times  \frac{\diff{^3 \3{p}_{1} } }{(2 \pi)^3 2 E_1} \times  \frac{\diff{^3 \3{p}_2 } }{(2 \pi)^3 2 E_2} \times \cdots \times  \frac{\diff{^3 \3{p}_N } }{(2 \pi)^3 2 E_N} \times \no 
&{} \times (2 \pi)^4 \delta^{(4)} \parz{k_A + k_B - \sum_{i=1}^N p_i} \, , \label{e.crossdef}
\end{align}
with the triangle function
$$
\lambda(s,m_A^2,m_B^2) \equiv s^2 + m_A^4 + m_B^4 - 2 s m_A^2 - 2 s m_B^2 - 2 m_A^2 m_B^2 \, .
$$
The total DIS cross section is
\begin{equation}
\label{e.sidis1}
E^\prime \frac{\diff{\sigma^{\rm tot}}{} }{\diff{^3 {\bf l}^\prime}{}} = \frac{2 \, \alpha_{\rm em}^2 }{\parz{s - \pmass^2} Q^4} \; 
L_{\mu \nu} W_{\rm tot}^{\mu \nu} \, .
\end{equation}
This fixes the normalization convention for the hadron and lepton
tensor combination $L_{\mu \nu} W_{\rm tot}^{\mu \nu}$, where the
"${\rm tot}$"-subindex indicates that this is  conventional DIS: totally inclusive in all
final state hadrons. Recall \fref{diagram} and \eref{theproc} for our
momentum labeling. The leptonic tensor is defined in the usual way:
\begin{equation}
\label{eq:lepttensor}
L_{\mu \nu} = 2 (l_\mu l^\prime_\nu + l^\prime_\mu l_\nu - g_{\mu \nu} \; l \cdot l^\prime) \, .
\end{equation}
So, the totally inclusive hadronic tensor is
\begin{align}
W^{\mu \nu}_{\rm tot} (P,q)
&{} \equiv 4 \pi^3 \sum_X \delta^{(4)}(P + q - P_X) \, 
\langle P, S | j^{\mu}(0) | X \rangle \langle X | j^{\nu}(0) | P, S \rangle \, .
\label{e.hadronictensor}
\end{align}
Here, the $\sum_X$ symbol is a sum over all possible final states $| X
\rangle$, including invariant integrals, 
$$
\int \frac{\diff{^3 \3{p}_i}{}}{2 E_{p_i} (2 \pi)^3} \, \cdots 
$$
over the momentum of each final state particle $p_i$. The incoming
hadron has a polarization specified by $S$. 

The SIDIS cross section is differential in the momentum of one observed final
state hadron of type $B$:
\begin{equation}
4 {\hadpsc}^0 E^\prime \frac{\diff{\sigma}{}}{\diff{^3 \mom{\bf l}{}{'}} \, \diff{^3 \mom{\bf P}{{\rm B}}{}}} 
= \frac{2 \, \alpha_{\rm em}^2 }{(s-\pmass^2) Q^4} \; 
L_{\mu \nu} W_{\rm SIDIS}^{\mu \nu}\, .
\label{e.sidis12}
\end{equation}
This fixes our normalization conventions for SIDIS, and gives a SIDIS hadronic
tensor:
\begin{equation}
\label{e.hadronictensortmd}
W^{\mu \nu}_{\rm SIDIS}(P,q,\hadpsc) 
\equiv \sum_X \delta^{(4)}(P + q - \hadpsc - P_X) \, 
\langle P, S | j^{\mu}(0) | \hadpsc,X \rangle \langle \hadpsc,X | j^{\nu}(0) | P,S \rangle \, .
\end{equation}
The same meaning applies to $\sum_X$ as in the totally inclusive case.
$| \hadpsc,X \rangle$ is a final state with at least one identified
hadron of type $B$. The sum over $X$ includes a sum over any number of
other final state particles, including other type-$B$ hadrons. Each
separate type-$B$ hadron in an event is counted, in accordance with the
definition of an inclusive cross section.

$W^{\mu \nu}_{\rm tot} (P,q)$ and $W^{\mu \nu}_{\rm
SIDIS}(P,q,\hadpsc)$ are the most convenient objects to work with
theoretically because they are Lorentz tensors directly related to
hadronic matrix elements of the electromagnetic current operator, and they are defined without reference to conventions associated with
choices of reference frames etc, so we will organize our structure
function analysis around them.

The relationship between the semi-inclusive and totally inclusive
cross sections follows from the definition in \eref{crossdef} (see,
e.g.,~\cite[Chapt. VII]{Byckling:1971vca}):
\begin{equation}
\sum_B \int \diff{^3 {\bf P}_B}{} \;\; \frac{\diff{\sigma}{}}{\diff{^3 {\bf P}_B}{} } 
= \langle N \rangle \sigma^{\rm tot} \, ,
\end{equation}
where $\langle N \rangle$ is the total average particle multiplicity, and
the sum is over all particle types. Thus,
\begin{equation}
 \sum_B \int \frac{\diff{^2 {\hadmomt{\rm b}}} 
 \diff{\hadmomplain^{z}}}{4 \hadmomplain^{0}}  W_{\rm SIDIS}^{\mu \nu} 
 = \sum_B 
 \int \frac{\diff{^2 {\hadmomt{\rm b}}} \diff{\zex}}{4 \zex}  W_{\rm SIDIS}^{\mu \nu} 
 = \langle N \rangle W_{\rm tot}^{\mu \nu} \, . \label{e.reduced2}
\end{equation}
Note that the integration measure in \eref{reduced2} is Lorentz
invariant, although we will continue to specify a photon frame for the
components, both for definiteness and because $\zex$ is defined in
terms of a photon frame momentum fraction.

\newpage

\noindent The usual structure function decompositions on $W^{\mu \nu}_{\rm tot}$ and 
$W^{\mu \nu}_{\rm SIDIS}$ are
\begin{align}
W^{\mu \nu}_{\rm tot} 
&{}= \left(-g^{\mu \nu} + \frac{q^{\mu} q^{\nu}}{q^2} \right) F_1^{\rm tot}(\xbj,Q^2) 
+ \frac{\left(P^\mu - q^\mu \frac{P \cdot q }{ q^2}\right) \left(P^\nu - q^\nu \frac{P \cdot q }{ q^2}\right)}{P \cdot q} F_2^{\rm tot}(\xbj,Q^2) \, \nonumber \\
&{}+ \text{Pol. Dep.} \, ,  \label{e.incstructdec} \\
W^{\mu \nu}_{\rm SIDIS} 
 &{}= \left(-g^{\mu \nu} + \frac{q^{\mu} q^{\nu}}{q^2} \right) 
 F_1(\xbj,Q^2,\zh,{\hadmomt{\rm b}}) 
  \, \nonumber \\
&{}
 +
\frac{\left(P^\mu - q^\mu \frac{P \cdot q }{ q^2}\right) \left(P^\nu - q^\nu \frac{P \cdot q }{ q^2}\right)}{P \cdot q} F_2(\xbj,Q^2,\zh,{\hadmomt{\rm b}}) \, 
+ \text{Pol. Dep.}\, .  \label{e.structdec}
\end{align}
``$\text{Pol. Dep.}$'' is a place holder for polarization and
azimuthal angle dependent terms, which we leave unspecified for now.
The structure functions' explicit dependence on $\pmass$ and $\hadmass$
has been dropped for brevity. While $\xbj$ and $\zh$ are shown as the
independent variables for the structure functions, it is useful to
view them as being themselves functions of $\xn$, $\zex$, $\pmass$ and
$\hadmass$. We have not done this here in order to avoid
over-complicating notations, but it is useful for making kinematical
approximations clear, as discussed in~\cite{Moffat:2019qll}.  The
differential SIDIS cross section in the Breit frame (or any photon
frame) is then
\begin{align}
\label{e.unpolstruct2}
& \frac{\diff{\sigma}{} }{\diff{\xbj}{} \diff{y}{} \diff{\psi}{} \diff{\zex}{} \diff{^2 \hadmomt{\rm b}{}} } = \frac{\alpha_{\rm em}^2 y}{4 Q^4 \zex } \; 
L_{\mu \nu} W_{\rm SIDIS}^{\mu \nu} \nonumber \\
& = 
\frac{\alpha_{\rm em}^2}{2 \xbj y \zex Q^2} \left[ \left( 1 - y - \frac{\xbj^2 y^2 M^2}{Q^2}\right) F_2 
+ y^2 \xbj F_1 \; + \text{Pol. Dep.} \right] 
\nonumber \\ 
& =  
\frac{\alpha_{\rm em}^2}{4 \xbj \zex y  Q^2} \left[ \left(1 + (1 -y)^2 + \frac{2 \xbj^2 y^2 M^2}{Q^2} \right) F_2 
- y^2 F_L \; + \text{Pol. Dep.} \right] \nonumber \\
&= \frac{\alpha_{\rm em}^2 y}{4 \xbj \zex Q^2 (1 - \kinep)} \bigl[  F_T + \kinep F_L \; + \text{Pol. Dep.} \bigr] \, .
\end{align}
In the last two lines
\begin{align}
F_T &\equiv 2 \xbj F_1 \, ,\label{e.FTdef} \\
F_L &\equiv \left(1 + \frac{4 \pmass^2 \xbj^2}{Q^2} \right) F_2 - 2 \xbj F_1 =  \left(1 + \frac{4 \pmass^2 \xbj^2}{Q^2} \right) F_2 - F_T\, , \label{e.FLdef}
\end{align} 
which are definitions generalized from the inclusive case to SIDIS. To
match with other common notational conventions, we have used
\begin{equation}
\gamma \equiv \frac{2 \pmass \xbj}{Q}\, , \qquad \kinep \equiv \frac{1 - y - \frac{\gamma^2 y^2}{4} }{1 - y + \frac{y^2}{2} + \frac{\gamma^2 y^2}{4} } \, , 
\end{equation}
along with the identities (see \cite[Eqs.~(2.8-2.13)]{Bacchetta:2006tn}),
\begin{equation}
\frac{1 - y + \frac{y^2}{2} + \frac{y^2 \gamma^2}{4} }{1 + \gamma^2} = \frac{y^2}{2(1 - \kinep)}\, ,
\qquad \frac{1 - y - \frac{y^2 \gamma^2}{4} }{1 + \gamma^2} = \frac{y^2 \kinep}{2(1 - \kinep)} \, .
\end{equation}
After changing variables from $\zex$ to $\zh$ \eref{unpolstruct2} becomes
\begin{align}
\frac{\diff{\sigma}{} }{\diff{\xbj}{} \diff{y}{} \diff{\psi}{} \diff{\zh}{}  \diff{^2 {\hadmomt{b}{}} }{}  } 
&{}=
\frac{\alpha_{\rm em}^2 y}{4 \zh \xbj Q^2 (1 - \kinep)} 
\frac{1}{\sqrt{1-\frac{4 \pmass^2 \xbj^2 M_{\rm B, T}^2}{Q^4 \zh^2}}} \bigl[  F_T + \kinep F_L  + \text{Pol. Dep.} \bigr] \, .
\label{e.unpolstruct4}
\end{align}
This Jacobian factor can be expressed in terms of combinations of $\zex$ and $\zh$, but we keep
the square root factors explicit to highlight the dependence on
transverse momentum via $M_{\rm B,T}^2$ at fixed $\zh$.

A convenient recipe for calculating structure functions is to contract
with Lorentz covariant extraction tensors, 
$\contractortot{\Gamma}^{\mu\nu}$, defined as
\begin{align}
\contractortot{g}^{\mu\nu} = g^{\mu\nu} ,\;\quad
\contractortot{PP}^{\mu\nu} = P^\mu P^\nu \, .
\label{e.PgPPtot}
\end{align}
Then
\begin{equation} 
F_1(\xbj,Q^2,\zh,{\hadmomt{\rm b}}) = \contractortot{1}^{\mu \nu} 
	W_{\mu \nu \, , {\rm SIDIS}} \, \qquad 
F_2(\xbj,Q^2,\zh,{\hadmomt{\rm b}}) = \contractortot{2}^{\mu \nu} 
	W_{\mu \nu \, , {\rm SIDIS}} \,  \label{e.contractorstot} \, ,
\end{equation}
where
\begin{align}
\contractortot{1}^{\mu\nu} {}& \equiv 
 -\frac{1}{2} {\rm P}_g^{\mu\nu}
   +\frac{2 Q^2 \xn^2}{(\pmass^2 \xn^2 + Q^2)^2} {\rm P}_{PP}^{\mu\nu}
			= -\frac{1}{2} \contractortot{g}^{\mu\nu} 
                          +\frac{2 \xbj^2}{Q^2} \contractortot{PP}^{\mu\nu} + O\Bigl( \frac{m^2}{Q^2} \Bigr) \, , \label{e.Con1}
                          \\
\contractortot{2}^{\mu\nu} {}& \equiv \frac{12 Q^4 \xn^3 \left(Q^2-\pmass^2 \xn^2\right)}
	 {\left(Q^2 + \pmass^2 \xn^2\right)^4}
    \left( {\rm P}_{PP}^{\mu\nu}
	 -\frac{\left(\pmass^2 \xn^2+Q^2\right)^2}{12 Q^2 \xn^2}
	   {\rm P}_g^{\mu\nu}\right) \nonumber \\ 
	   & = \frac{12  \xbj^3}{Q^2} \contractortot{PP}^{\mu\nu} 
                          - \xbj \contractor{g}^{\mu\nu} + O\Bigl( \frac{m^2}{Q^2} \Bigr) \, . \label{e.Con2}
\end{align}
From \eref{reduced2}
\begin{align}
\sum_B \int \, \frac{\diff{\zex} \, \diff{^2 {\hadmomt{\rm b}}}}{4 \zex} \, F_{j}(\xbj,Q^2,\zh,{\hadmomt{\rm b}}) 
&  = \langle N \rangle F_{j}^{\rm tot}(\xbj,Q^2) \, , \label{e.inclred}
\end{align} 
where $j$ labels a structure function, i.e., $j \in \{1,2,T,L\}$.
To reproduce the equations of Ref.~\cite{Bacchetta:2006tn}, we have define barred 
structure functions: 
\begin{equation}
\bar{F}_{j} 
= 
\frac{1}{4 \zh \left( 1 + \frac{\gamma^2}{2 \xbj} \right)} \frac{F_{j}}{\sqrt{1-\frac{4 \pmass^2 \xbj^2 M_{\rm B, T}^2}{Q^4 \zh^2}}} \, .
\label{e.barred}
\end{equation} 
Substitute into~\eref{unpolstruct4} to get
\begin{align}
\label{e.unpolstruct5}
\frac{\diff{\sigma}{} }{\diff{\xbj}{} \diff{y}{} \diff{\psi}{} \diff{\zh}{}  \diff{^2 {\hadmomt{\rm b}{}} }{}  }  =
\frac{\alpha_{\rm em}^2 y}{\xbj Q^2 (1 - \kinep)} 
\left( 1 + \frac{\gamma^2}{2 \xbj} \right) \bigl[  \bar{F}_T + \kinep \bar{F}_L + \text{Pol. Dep.} \bigr] \, .
\end{align}
Now~\cite[Eq.~(2.7)]{Bacchetta:2006tn} can be used to fill in the remaining polarization 
and $\phi$-dependent structure functions. The barred normalization convention 
in \eref{barred} is defined so that structure functions exactly obey a particularly 
convenient energy sum rule~(\aref{bdgmms}) found in \cite[Eqs.~(2.18-2.21)]{Bacchetta:2006tn}:
\begin{equation}
\sum_{B} \int \diff{\zh}{} \diff{^2 {\hadmomt{\rm b}{}} }{} \zh \bar{F}_{j} = F^{\rm tot}_{j}   \, . \label{e.esumrule}
\end{equation}
Note that a factor of hadron multiplicity does not appear.
Equation~\eqref{e.inclred} becomes
\begin{equation}
\sum_{\rm B} \int \, \diff{\zh} \, \diff{^2 {\hadmomt{\rm b}{}}} \, \bar{F}_{j}
= \langle N \rangle  \left( 1 + \frac{\gamma^2}{2 \xbj} \right)^{-1} F_{j}^{\rm tot} \label{e.reduc1f} \, .
\end{equation}
The forms of \erefs{esumrule}{reduc1f} are only valid if the barred normalization 
conventions from \eref{barred} are used for the structure function normalizations, making 
them very useful for some practical applications. An advantage of the unbarred 
convention is that the structure functions have a direct connection to the matrix 
elements of current operators via \eref{hadronictensortmd}, and their Lorentz covariant 
structure function decomposition with the standard normalization conventions in \eref{sidis12} and \eref{structdec}.

The unobserved invariant mass-squared in inclusive DIS is 
\begin{equation}
W_{\rm tot}^2 = \pmass^2 + \frac{Q^2 (1-\xbj)}{\xbj} \, .
\end{equation}
In SIDIS it is
\begin{align}
W_{\rm SIDIS}^2 &{}= \pmass^2 + \hadmass^2 + \frac{Q^2 (1 - \xbj - \zh)}{\xbj} 
+ \frac{Q^4 \zh \parz{\sqrt{1 + \frac{4 \pmass^2 \xbj^2}{Q^2}} \sqrt{1 - \frac{4 \pmass^2 \xbj^2 M_{\rm B, T}^2}{\zh^2 Q^4}} - 1} }{2 \pmass^2 \xbj^2} \,  \no 
   &{} \stackrel{\pmass, \hadmass \to 0}{=} 
   \frac{Q^2 (1-\xbj) (1-\zh)}{\xbj}-\frac{\hadmomtsc{2}{}}{\zh} \, . \label{e.Wsidis}
\end{align}
Note that if both $\zh$ and $\xbj$ are close to 1, then
$|\hadmomtsc{}{}|$ cannot be much greater than zero without hitting
the resonance region of $W_{\rm SIDIS}^2 \approx 0$.

\section{Purely Kinematical Approximations}
\label{s.kins}
Since we have not discussed the theory underlying the structure
functions, all small mass approximations mentioned so far are
unambiguously kinematical. For example, the usual $\xn \approx \xbj$
and $\zex \approx \zh$ follow from expanding in $\xbj^2 \pmass^2/Q^2$:
\begin{align}
\xn  &{}= \xbj \left[ 1 - \frac{\xbj^2 \pmass^2}{Q^2} 
+ \order{\frac{\xbj^4 \pmass^4}{Q^4} }  \right] \, , \label{e.xnexp} \\
\zex  &{}= \zh \left[ 1 - \frac{\xbj^2 \pmass^2}{Q^2} 
\left( 1 + \frac{\hadmomtsc{2}{\rm b,}}{\zh^2 Q^2} \right)  
        + \left( \frac{\xbj^2 \pmass^2}{Q^2} \right)^2 
        \left(\frac{\hadmomtsc{2}{\rm b,}}{\zh^2 Q^2} 
        - \frac{\hadmomtsc{4}{\rm b,}}{\zh^4 Q^4} 
        + 2 -\frac{\hadmass^2}{\zh^2 \pmass^2 \xbj^2} \right)\right.
        \,  \nonumber \\
        &{}
        \left. + {}\,\order{ \frac{\xbj^6 \pmass^6}{Q^6}  } \right] \, . \label{e.znexp}
\end{align}
If hadron masses are neglected, then $P$ and $\hadpsc$
become the approximate $\tilde{P}$ and $\tilde{\hadpsc}$, which we define as
\begin{align}
\tilde{\Pmom{}{\rm b}} &{}= \parz{\frac{Q}{\xbj \sqrt{2}}, 0, \T{0}{} } \, , \\
\hadmomap{}{\rm b}  &{}= \zh \parz{\frac{\qmomtsc{2}{} }{\sqrt{2} Q}, \frac{Q}{\sqrt{2}}  ,
 - \qmomt{} } \, ,
\end{align}
that is, \eref{pmombreit} and \eref{hadronpara} but with all hadron
masses set equal to zero. In the most common treatments of SIDIS, $P$
and $\hadpsc$ in \erefs{incstructdec}{structdec} are replaced with
$\tilde{P}$ and $\hadmomap{}{}$, and $\xn$ and $\zex$ are replaced
with $\xbj$ and $\zh$ inside the structure functions, which is a good
approximation in the $m/Q \to 0$ limit as long as the structure
functions are reasonably smooth functions of $\xn$ and $\zex$. In
\cite{Moffat:2019qll} that was called the massless target approximation
(MTA) for inclusive DIS, and an obvious extension applies to SIDIS.
In that case, \erefs{incstructdec}{structdec} become
\begin{align}
\tilde{W}^{\mu \nu}_{\rm tot} 
&{}= \left(-g^{\mu \nu} + \frac{q^{\mu} q^{\nu}}{q^2} \right) \mathcal{F}_1^{\rm tot}(\xbj,Q^2) 
+ \frac{\left(\tilde{P}^\mu - q^\mu \frac{\tilde{P} \cdot q}{q^2}\right) \left(\tilde{P}^\nu - q^\nu \frac{\tilde{P} \cdot q}{q^2}\right)}{\tilde{P} \cdot q} \mathcal{F}_2^{\rm tot}(\xbj,Q^2) \; \nonumber \\
&{}+ \text{Pol. Dep.} \, ,  \label{e.incstructdecMHA} \\
\tilde{W}^{\mu \nu}_{\rm SIDIS} 
 &{}= \left(-g^{\mu \nu} + \frac{q^{\mu} q^{\nu}}{q^2} \right) \mathcal{F}_1(\xbj,Q^2,\zh,{\hadmomtap{\gamma}})
 \, \nonumber \\
&{} +
\frac{\left(\tilde{P}^\mu - q^\mu \frac{\tilde{P} \cdot q}{q^2}\right) \left(\tilde{P}^\nu - q^\nu \frac{\tilde{P} \cdot q}{q^2}\right)}{\tilde{P} \cdot q} \mathcal{F}_2(\xbj,Q^2,\zh,{\hadmomtap{\gamma}}) 
+ \text{Pol. Dep.}\, .  \label{e.structdecMHA}
\end{align}
Extracting the structure functions in
\erefs{incstructdecMHA}{structdecMHA} requires,  instead of
\erefs{Con1}{Con2},
\begin{equation}
\contractortotap{1}^{\mu\nu} = -\frac{1}{2} \contractortot{g}^{\mu\nu} 
                          +\frac{2 \xbj^2}{Q^2} \contractortotap{PP}^{\mu\nu}  
                          \, , \qquad
\contractortotap{2}^{\mu\nu} = \frac{12  \xbj^3}{Q^2} \contractortotap{PP}^{\mu\nu} 
                          - \xbj \contractor{g}^{\mu\nu}
                          \, ,
\end{equation}
where $\contractortotap{PP}^{\mu\nu} = \tilde{P}^{\mu}
\tilde{P}^{\nu}$.  Our~\eref{incstructdecMHA} is Eq.~(18) from
\cite{Moffat:2019qll}, with the calligraphic notation explained there.
Equation \eqref{e.structdecMHA} is the analogous approximation for the
SIDIS cross section. The MTA greatly simplifies kinematical relations
at large $Q$. 

The ratios
\begin{equation}
\frac{\xn}{\xbj} \, , \qquad \frac{\zex}{\zh} \,  \label{e.xzrats}
\end{equation}
are measures of the quality of the MTA. They must not deviate too much from $1$ if 
the standard massless approximations are  to be considered valid. (See 
\sref{examples} for some examples.) 

This exhausts the approximations that can be assessed
entirely independently of questions about the partonic dynamics responsible for
the behavior of the structure functions themselves. 

Refs.~\cite{Accardi:2014qda,Guerrero:2015wha} made first attempts 
to incorporate kinematical improvements to collinear 
QCD factorization by keeping $\hadmass$ in kinematical 
factors. They point out the importance of this for moderate-to-low $Q$ SIDIS. However, 
they explicitly drop $\hadmomtsc{2}{\rm b,}$-dependence in an attempt to stay within a collinear factorization framework.
Note from \eref{znexp}, however, that 
it is not consistent with collinear factorization power counting
 to simultaneously retain $\pmass$ and $\hadmass$ dependent 
kinematical power corrections while neglecting $\hadmomtsc{2}{\rm b,}$ dependent 
corrections, even for $\hadmomtsc{2}{\rm b,} \sim m^2$.
The first non-vanishing $\hadmass$-dependent correction
term
\begin{equation}
\frac{\hadmass^2}{\zh^2 \pmass^2 \xbj^2} \left( \frac{\xbj^2 \pmass^2}{Q^2} \right)^2 \label{e.terma}
\end{equation}
is the same size as the 
\begin{equation}
\frac{\hadmomtsc{2}{\rm b,}}{\zh^2 Q^2} \left( \frac{\xbj^2 \pmass^2}{Q^2} \right) \label{e.termb}
\end{equation}
term when $\hadmomtsc{2}{\rm b,}$ is small. 
And, \eref{termb} is actually the dominant power-correction term when 
$\hadmomtsc{2}{\rm b,}/\zh^2$ approaches order $Q$. 
The difficulty is that collinear factorization methods only characterize dependence on light-cone momentum fractions of the 
final state 
hadron, like $\zex$, with only $\zh$, $Q$, and $\xbj$ known. This is not a problem if keeping only the first 
term in the expansion on the right of \eref{znexp} is valid.
But the \emph{exact} $\zex$ requires knowledge not just of $\hadmass$ and $\pmass$, but also of (both small and large) $\hadmomtsc{2}{\rm b,}/\zh^2$. 
So if it turns out that final state mass effects are large enough that they have to be accounted for, then it must be done in combination with an account of small transverse momentum dependence effects (e.g., TMD factorization), not independently of it.

\section{Partons}
\label{s.partons}

So far, we have only discussed definitions and relativistic
kinematics, with no mention at all of partons or dynamics. The
question now is the following: Assuming that the configuration of initial and final 
hadrons
is the result of scattering and fragmentation by small-mass
constituents (i.e., partons), what are the possible kinematical
configurations of those constituents, given a set of assumptions about
their intrinsic properties?  For now, we do not necessarily
identify these partons with a particular theoretical approach
or even real QCD, though ultimately we have that in mind.
\begin{figure*}
\centering
\includegraphics[scale=0.6]{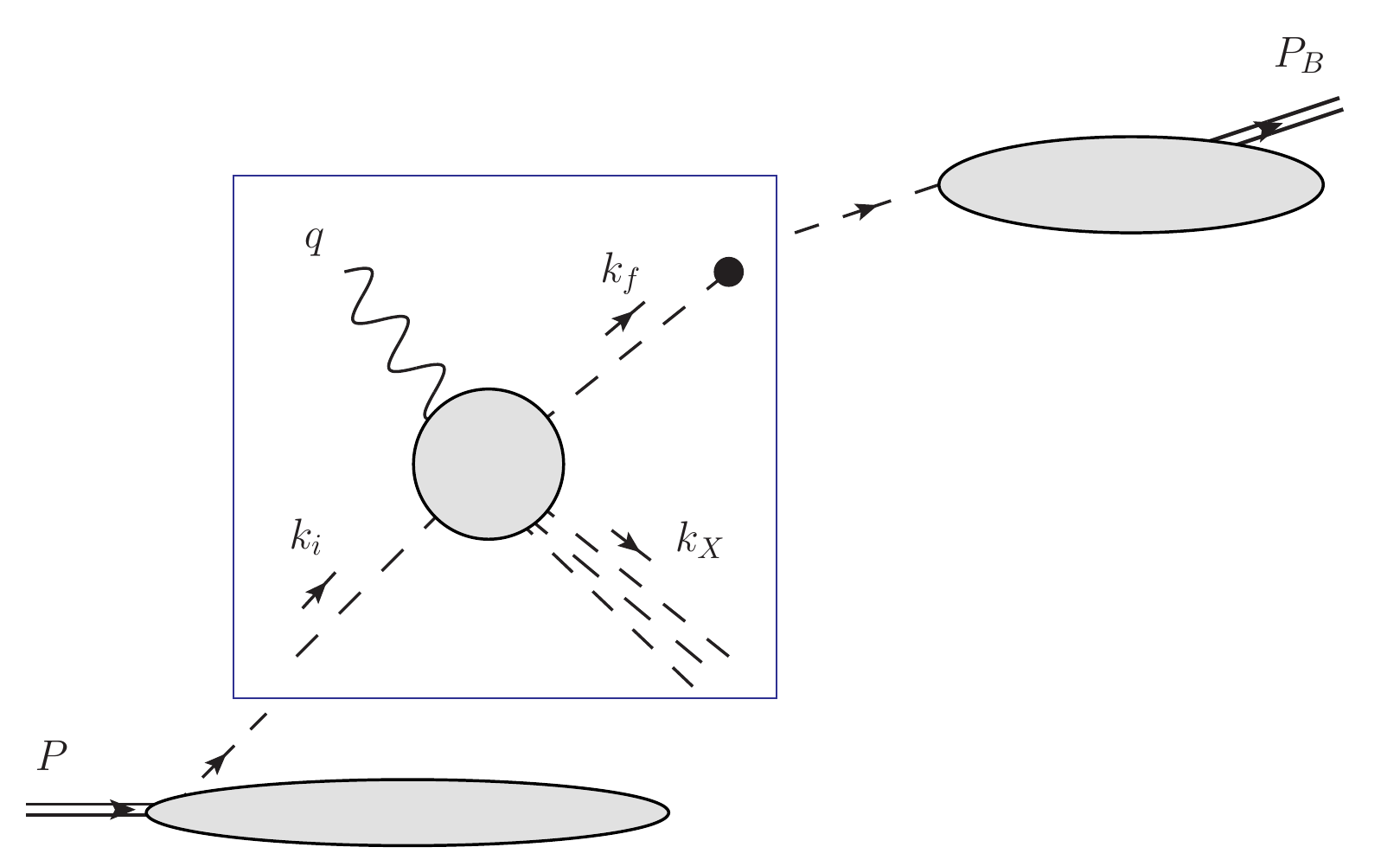}
\caption{Momentum labeling in the partonic subprocess.}
\label{f.partonic}
\end{figure*}

This kind of very general partonic picture is illustrated
in~\fref{partonic}. We start by exploring the possibility that
the produced hadron is collinear to an outgoing parton (a ``current''
hadron). We need clear steps for asking how reasonable it is to assume
that a given external kinematical configuration for measured hadrons
maps to current region partonic kinematics. The incoming hadron and
its remnants are represented by the lower blob while the final state
hadron emerges from a final state blob at the top of the diagram. 
Dashed lines represent the flow of momentum. It is very important for the 
discussion below to understand 
that they do not necessarily represent single quarks or gluons, and in reality they 
may correspond to groups of particles. What is important for us is only the flow of 
four-momentum through the process. Moreover, it is assumed that the momenta 
of these lines is known exactly and are never approximated. Although the word
``parton'' often implies a massless on-shell approximation for single particle lines, to  
keep language 
 reasonably simple, we will nevertheless continue to call these dashed lines "partons."
 The picture 
in \fref{partonic} \emph{does} imply that 
quantities like $|\initq^2|$ and $|\finalq^2|$ are small, and much of the discussion in 
this section will be about addressing the question of what is meant by ``small.'' So to summarize, ``partonic" dashed lines 
represent the flow of a momentum with small invariant energy. In practical situations, they will often turn out to refer to actual quark and/or gluon lines, but they do not need to generally.

The
partonic subprocess in~\fref{partonic}  is marked off in a blue box. 
A black dot indicates
the parton we associate with an observed hadron. The momentum $\initq$
is the incoming struck parton momentum, and there is at least one
hadronizing parton $\finalq$. The $\xmom$ momentum labels the total
momentum of all other unobserved partons combined. Outside the box in
\fref{partonic}, the position of the hadron implies a current region
picture, though an analogous picture of course applies to the target
region case. We ask questions about partonic regions in the context of
the steps needed to factorize graphical structure in a manner
consistent with particular partonic pictures. Our general view  
of factorization is based on that of
Collins~\cite{Collins:2011qcdbook,Collins:1988gx} and
collaborators, though the same statements apply to most other approaches. 

We are interested in the kinematics of the 
$k_{\rm i} + q \to k_{\rm f} + k_X$ subprocess 
and how closely it matches the overall $P + q \to \hadpsc + X$
process under very general assumptions. Specific realizations of
the partonic subprocess, each of which can contribute to a different
kinematical region, are shown in \fref{basickinematics}.
\begin{figure}
\centering
\centering
  \begin{tabular}{c@{\hspace*{5mm}}c}
    \multicolumn{2}{c}{
    \includegraphics[scale=0.5]{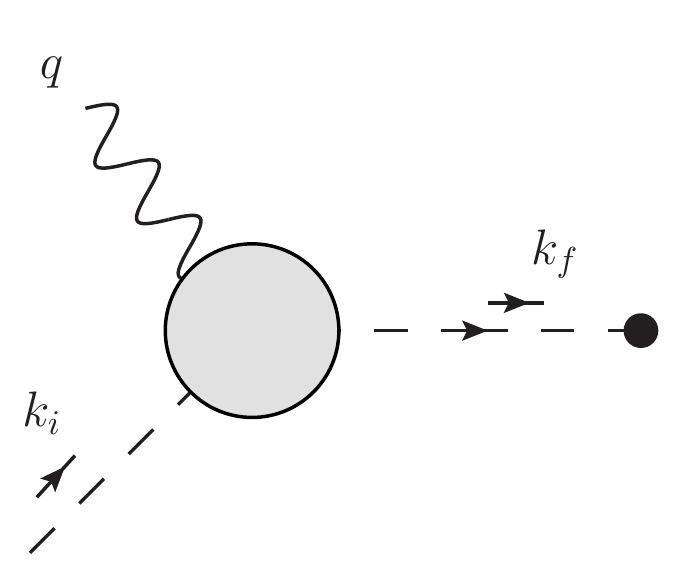}
  }
  \\
  \multicolumn{2}{c}{(a)} \\ \\
    \includegraphics[scale=0.5]{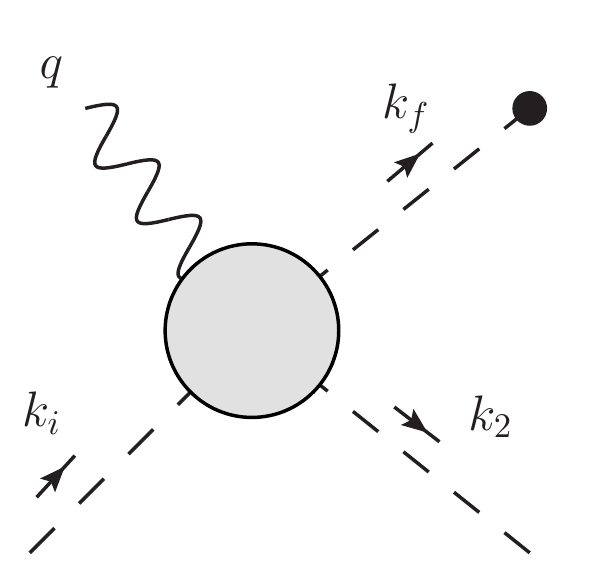}
    \hspace{0.5cm}
    &
    \hspace{0.5cm}
    \includegraphics[scale=0.5]{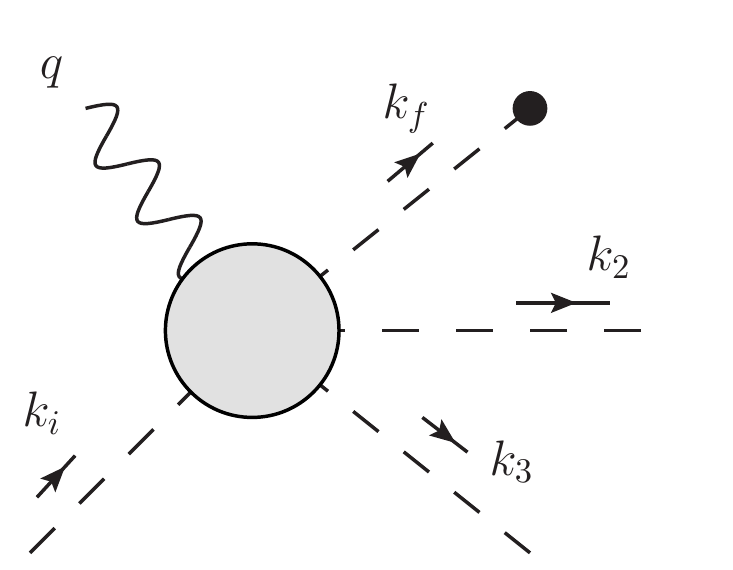}
  \\
  (b) & (c) \\ \hspace{0.5cm}
  \end{tabular}
\caption{
  Examples of hard kinematics. Graph (a) represents handbag kinematics.
  Graph (b) is $2 \to 2$ kinematics, which can represent, for instance, the first non-vanishing contribution when we specialize to massless pQCD graphs at large transverse momentum. Graph (c) is $2 \to 3$ kinematics. We remark that in general, in Graphs (a), (b) and(c) the dashed lines may represent groups of particles, such as those making up a gauge link.}
\label{f.basickinematics}
\end{figure}
We will analyze the subprocess in the Breit frame and write
\begin{equation}
\initq^{\rm b} = 
\parz{\frac{Q}{\xnp \sqrt{2}}, \frac{\xnp (\initq^2 + \initqtb^2)  }{\sqrt{2} Q}, \initqtb } \, ,
\qquad 
 \finalq^{\rm b} = \parz{\frac{\finalqtb^2 + \finalq^2 }{\sqrt{2} \zexp Q}, 
 \frac{\zexp Q}{\sqrt{2}}  , \finalqtb } \, . \label{e.parlabels2}
\end{equation}
Hats always indicate a partonic kinematical variable, whereas $\xi$
and $\zeta$ are momentum fractions (see below). We will write the
transverse momentum as
\begin{equation}
\finalqtb = - \zexp \qmomt{}{} + \delta \T{k}{} \, . \label{e.finalz}
\end{equation}
In the hadron frame, \eref{Vtrans} gives
\begin{equation}
\finalqtH = \delta \T{k}{} + \text{Power Suppressed} \, ,
\end{equation}
so $\delta \T{k}{}$ is good for characterizing an intrinsic relative
transverse momentum in the large $Q$ limit; in \eref{parlabels2} intrinsic transverse
momentum is $\delta \T{k}{}$ when $\Tsc{q}{} = 0$.  For nearly
on-shell partons, 
\begin{equation}
 |\initq^2|,\, |\finalq^2| = \order{m^2} \, .
\end{equation}
In the limit where $m \ll Q$ and $\xbj$, $\zh$, $\Tsc{q}{}$ are fixed,
the outgoing parton is exactly aligned with the observed hadron so
long as 
\begin{equation}
\delta \Tscsq{k}{} = \order{m^2} \, .
\end{equation}
We have defined the Breit frame momentum fractions and Breit frame
$\xnp$, $\zexp$ analogous to $\xn$ and $\xbj$:
\begin{equation}
\initq^+ \equiv \xi \Pmom{+}{\rm b} \, , \qquad \hadmom{-}{\rm b} \equiv \zeta \finalq^- \, , \qquad \xnp \equiv  -\frac{\qmom{+}{\rm b}}{\initqb^+} = \frac{\xn}{\xi} \, 
, \qquad \zexp \equiv  \frac{\finalqb^-}{\qmom{-}{\rm b}} = \frac{\zex}{\zeta} \, .
\end{equation}
For fixed $\xnp$, $\zexp$ and $\Tscsq{q}{}$, $\xmom^2$ is calculable from momentum conservation,
\begin{equation}
\xmom^2 = \left(\initq + q - \finalq \right)^2 \, . \label{e.xmomva}
\end{equation}
It will also be useful to define a momentum variable
\begin{equation}
k \equiv \finalq - q \, . \label{e.kdef}
\end{equation}
It is sometimes useful to have $k$ in terms of $k_X^2$ instead of
$\zexp$.  For example, in the special case that 
$\initq^2 = \finalq^2 =  \initqtb^2 =  \delta \T{k}{}^2 = 0$
\begin{align}
k_{\rm b}^+ 
&{}=\frac{Q}{\sqrt{2}} \parz{1 +\frac{\Tscsq{q}{}}{Q^2} 
\parz{\frac{1 - \xnp (1 + \xmom^2/Q^2)}{1 - \xnp (1 - \Tscsq{q}{}/Q^2)}} } 
   = \frac{Q}{\sqrt{2}} \parz{1 + \frac{\Tscsq{q}{}}{Q^2} + \cdots } \, , \label{e.kpapp} \\
k_{\rm b}^- 
&{}= -\frac{Q}{\sqrt{2}} 
\parz{1 - \frac{1 - \xnp (1 + \xmom^2/Q^2)}{1 - \xnp (1 - \Tscsq{q}{}/Q^2)}} 
   = -\frac{\xnp Q}{(1 - \xnp)\sqrt{2}} 
   \parz{\frac{\Tscsq{q}{}}{Q^2} + \frac{\xmom^2}{Q^2} + \cdots }  \, , \\
\T{k}{} &{}= - \T{q}{} \parz{\frac{1 - \xnp (1 + \xmom^2/Q^2)}{1 - \xnp (1 - \Tscsq{q}{}/Q^2)}} \, 
           = - \T{q}{} \parz{1 - \frac{\xnp}{1 - \xnp} \parz{\frac{\Tscsq{q}{}}{Q^2} + \frac{\xmom^2}{Q^2}}  + \cdots }  \, . \label{e.kperpapp}
\end{align}
On the second line, the "$\cdots$" represents higher powers in
an expansion in small $\Tscsq{q}{}/Q^2$ and $\xmom^2/Q^2$.  When
$\Tscsq{q}{}/Q^2 \to 0$ and $\xmom^2/Q^2 \to 0$, the kinematics of the
struck parton approach the kinematics of TMD factorization, or the 
handbag contribution in collinear factorization, with the
errors in each component proportional to $\Tscsq{q}{}/Q^2$. 

The most basic of partonic approximations is that the masses and
off-shellness of partons is small relative to the hard scale:
\begin{equation}
\initq^2/Q^2 \to 0 \qquad \finalq^2/Q^2 \to 0 \, . \label{e.smallsq}
\end{equation}
On top of these, other approximations are normally needed. For
instance, in the current region $\finalq$ is
aligned with the final state hadron and
\begin{equation}
\finalq \cdot \hadmomplain{}{} \to 0 \, .
\end{equation}
Beyond these, still further approximations apply to different specific
partonic subprocesses.  First, in the $2 \to 1$ process of
\fref{basickinematics}(a), $\initq \to k$, and the $1/Q^2$-suppressed terms in equations like \erefs{kpapp}{kperpapp} are dropped. For a
hard $2 \to 2$ process shown in \fref{basickinematics}(b), 
$| k^2 |
\sim Q^2$ while $\xmom^2/Q^2 \to 0$. If both $|k^2|$ and $\xmom^2$ are
large, then at least three partons (e.g., \fref{basickinematics}(c))
are ejected at wide angles from the hard collision. 
For fixed $\xn$, $\zex$, $Q^2$, and $\hadmomplaint{}{}$, only certain
$\initq$ and $\finalq$ are consistent with any given picture in
\fref{basickinematics}. 

For example, say we wish to interpret a particular SIDIS region with a
partonic configuration like \fref{basickinematics}(a), corresponding
to the current fragmentation region. For a partonic description to
hold at all, a minimum requirement is that ratios like \eref{smallsq} are
very small. So define a ratio

\begin{equation}
\text{General Hardness Ratio} = \ratgen \equiv 
	\text{max} \left( \left| \frac{\initq^2}{Q^2} \right|, \left| \frac{\finalq^2}{Q^2} \right|, 
	\left| \frac{\delta \Tscsq{k}{}}{Q^2} \right|\right)  \, .
\label{e.R0}
\end{equation}
and consider regions of $Q$ where $\ratgen$ is less than a certain
numerical size for a given set of estimates for $\initq^2$ and
$\finalq^2$. Next, since scattering is assumed to be in the current
region in 
\fref{basickinematics}(a), the ratio
\begin{equation}
\text{Collinearity} = \ratcur \equiv \frac{\hadpsc \cdot \finalq}{\hadpsc \cdot \initq} \,  ,
\label{e.R1}
\end{equation}
must also be small. See Ref.~\cite{Boglione:2016bph} for more
discussion -- $\ratcur$ corresponds to $R$ from that reference.  The 
expression for $\ratcur$ in terms of the variables in
\eref{hadronpara} and \eref{parlabels2} is straightforward, but slightly cumbersome 
and not instructive, so we will not write it explicitly here.

The $2 \to 1$ partonic kinematics only apply if $k^2/Q^2 \approx 0$, an approximation that fails if transverse momentum is too large. So define  
another ratio,
\begin{equation}
\text{Transverse Hardness Ratio} = \ratTM \equiv \frac{|k^2|}{Q^2} \, . \label{e.ratTM}
\end{equation}
$\ratTM$ is small for $2 \to 1$ partonic kinematics.  From
\eref{parlabels2},
\begin{equation}
\ratTM = 
\left|-(1-\zexp) - \zexp \frac{\Tscsq{q}{}}{Q^2} - \frac{(1 - \zexp) \finalq^2}{Q^2 \zexp} - \frac{\delta \momt{k}{}{2} }{\zexp Q^2} + \frac{2 \T{q}{} \cdot \delta \T{k}{} }{Q^2} \right|
        \approx (1-\hat{z}_N) + \hat{z}_N \frac{\Tscsq{q}{}}{Q^2} \, . 
\label{e.R2}
\end{equation}  
 Note that this suggests $\Tsc{q}{}$ from \eref{qtratdef} as the most
useful transverse momentum for quantifying transverse momentum
hardness relative to $Q$; if $\Tscsq{q}{}/Q^2 \sim 1$, then $\ratTM
\sim 1$ for both large and small $\hat{z}_N$ while if $\Tscsq{q}{}/Q^2
\ll 1$ and $\zeta \sim \zex$ (as in the current fragmentation region
with TMDs) then $\ratTM \ll 1$ (see also discussion in Ref.~\cite{Gonzalez-Hernandez:2018ipj}).

If the SIDIS region corresponds to $2 \to 2$ hard partonic kinematics,
then $\ratTM$ must be large ($\sim 1$).  However, then the ratio
$\xmom^2/Q^2$ must be small since there is only one unobserved parton,
and its invariant mass must be small relative to hard scales to
qualify as a single massless parton. (See~\fref{basickinematics}(b).) If $k_2$ is a massless on-shell quark 
or gluon, then $k_2^2 =0$ and this places a strong kinematical constraint on relationship between the momentum fractions $\xi$ and $\zeta$. See, for 
example, Eq.(83) of \cite{Nadolsky:1999kb}.
So define one more ratio,
\begin{equation}
\text{Spectator Virtuality Ratio} =  \ratX \equiv  \frac{|\xmom^2|}{Q^2} \, . 
\label{e.R3}
\end{equation}
Large $\ratTM$, but small $\ratX$, corresponds to $2 \to 2$ parton
kinematics. Large $\ratTM$ \emph{and} large $\ratX$ corresponds to
partonic scattering with three or more final state partons, such as
\fref{basickinematics}(c). 

To see that the size of $\ratTM$, \eref{R2}, reflects the importance of
transverse momentum, we repeat an argument very similar to 
that on page 4 of \cite{Gonzalez-Hernandez:2018ipj}. Note that Feynman graphs corresponding to the
inside of the box in \fref{basickinematics} contain propagator
denominators of the form
\begin{equation}
\frac{1}{k^2 + \order{m^2}} \, , \qquad  \frac{1}{k^2 + \order{Q^2}} \, , \label{e.denomsizes}
\end{equation}
where the denominators with $+\order{Q^2}$ arise in corrections to the
virtual photon vertex or internal propagators from the emission of wide-angle
$\xmom$ partons.  Note also that $k \cdot q \sim q \cdot P =
\order{Q^2}$.   The possible approximations to these denominators
are representative of the approximations needed in derivations of
factorization.  If $|k^2| \sim Q^2$, the $\order{m^2}$ terms in the
denominators are negligible so that the part of the graph inside the box can be
calculated in perturbative QCD using both $Q^2$ and $k^2$ as equally
good hard scales. In this case, and $\xmom^2 \ll Q^2$, then \fref{basickinematics}(b) becomes the
relevant picture. However, if $|k^2| \ll Q^2$, the $\order{m^2}$ terms
in the first of the denominators in \eref{denomsizes} must be kept.
Then, a $|k^2|/Q^2 \ll 1$ approximation in the second denominator can
be used, and it is this type of approximation that leads to TMD
factorization at small transverse momentum. This is the handbag topology in
\fref{basickinematics}(a). Note that the $k$ line has become the
target parton.  Using \eref{parlabels2} and \eref{kdef} for $k^2$ gives \eref{R2}. 

In perturbative QCD, the lowest order (in $\order{\alpha_s}$) contribution to large transverse
momentum is the partonic $2 \to 2$ process. Again, all
partons are massless and on-shell, and the picture is
\fref{basickinematics}(b). Since there is only one unobserved massless
parton in this region, it correspond to $k_X^2 = 0$.  To see that it
is the ratio $\ratX$ in \eref{R3} that must be small in this region,
consider how the size of $k_X^2$ affects the denominators in
\eref{denomsizes} at fixed $\hat{x}_N$, large $\Tsc{q}{}$,
and $Q^2$ by expressing $|k^2/Q^2|$ in terms of $k_X^2$ instead of
$\hat{z}_N$:
\begin{align}
	\left|\frac{k^2}{Q^2} \right| = 
  \frac{1}{1-\hat{x}_N+\hat{x}_Nq^2_{\rm T}/Q^2}
  \left[
    \frac{q_{\rm T}^2}{Q^2} 
    + \hat{x}_N\frac{k_X^2}{Q^2}\left(1-\frac{q^2_{\rm T}}{Q^2}\right)
  \right] \, .
\label{e.ksq}
\end{align}
To get a simple form, we have already assumed here that $\initq^2$ and
$\finalq^2$ are negligible.  In propagators, therefore, the size of
$k^2$ is independent of $k_X^2$ at large $\Tscsq{k}{}$ if $k_X^2 /Q^2
\ll 1$ and $\hat{x}_N$ is not too close to $1$.  Otherwise, if $\ratX$ 
becomes large, the $2 \to 3$ or greater cases are
likely the more applicable partonic subprocesses. In pQCD this means
that $\order{\alpha_s^2}$ or higher calculations are needed.

Different combinations of sizes for the above ratios correspond to
other regions. For example, the target fragmentation region handles
cases where $\ratcur$ gets large -- see \sref{target} below.
All of the approximations discussed above are intertwined in potentially
complicated ways, especially when $Q$ is not especially large and mass
effects may be non-negligible. This can make even crude,
order-of-magnitude estimates of their effects nontrivial, although the
influence of model assumptions should diminish rapidly at large $Q$. The catalogue of ratios represented by the $\ratgen$-$\ratX$ is meant to make this more straightforward to check.

A choice concerning acceptable ranges of $\ratgen$, $\ratcur$,
$\ratTM$, and $\ratX$ translates into a choice about the range of
possible reasonable values for the components of $\initq$ and
$\finalq$.  In practice, this might be more conveniently stated in
reverse. That is, one starts with general expectations regarding the
sizes of the partonic components of $\initq$ and $\finalq$ based on
models and/or theoretical considerations. The question then becomes
whether the resulting $\ratgen$, $\ratcur$, $\ratTM$, and $\ratX$ are
consistent with a particular region of partonic kinematics (hard,
current region, large transverse momentum, etc). 

Our aim here is not to
address any particular theoretical framework for estimating intrinsic
properties of partons, or to estimate exactly acceptable ranges for
the above ratios, but only to demonstrate how, once these choices are
made, they fix the relationship between external kinematics and the
region of partonic kinematics. 

\section{Rapidity}
\label{s.rapidity}
It is often useful to express results in terms of rapidity instead of
$\zex$ or $\zh$.  In the Breit frame, 
\begin{equation}
\py \equiv \ln \parz{\frac{Q}{\xn \pmass}} \, , \qquad \hady \equiv \ln \parz{\frac{\hadtmass}{\zex Q}} \, . \label{e.hadronraps}
\end{equation}
The boost invariant rapidity difference is
\begin{equation}
\Delta y \equiv \py - \hady = \ln \parz{\frac{\zex Q^2}{\xn \pmass  \hadtmass }} \, .
\end{equation}
If $\xn \approx \zex$ and $\hadtmass \approx \pmass$, then
the produced hadron rapidity is approximately the negative of the
proton rapidity. For fixed $\zex/\xn$, fixed $\hadtmass$ and large
$Q$
\begin{equation}
e^{\Delta y} = \order{\frac{Q^2}{m^2}} \, , \qquad e^{-\Delta y} = \order{\frac{m^2}{Q^2}} \, .
\end{equation}
$\zh$ in terms of $\hady$ is~\cite{Boglione:2016bph}
\begin{equation}
\zh  = \frac{\xn \hadtmass \pmass}{Q^2 - \xn^2 \pmass^2} \parz{e^{\Delta y}  + e^{-\Delta y} } \approx \frac{\xbj \hadtmass \pmass}{Q^2} e^{\Delta y}   \, .
\end{equation}
In terms of $\zh$, the rapidity of the hadron in the Breit frame is double valued:
\begin{equation}
\hady^{\pm} =
   \ln \left[\frac{Q \zh \left(Q^2-\xn^2 M_p^2\right)}{2 \xn^2 \pmass^2
   \hadtmass } \pm \frac{Q}{\xn \pmass} \sqrt{\frac{\zh^2 \left(Q^2-\xn^2 \pmass^2\right)^2}{4 \xn^2 \pmass^2
   \hadtmass^2}- 1} \; \right] \, \approx \ln \parz{\frac{\hadtmass}{\zh Q}} \, .
\label{eq:rapvalue}
\end{equation}
The ``$+$'' solution corresponds to a hadron with large rapidity in the
direction of $P$, while the ``$-$'' solution corresponds to a
rapidity in the opposite direction, and thus is more consistent with
current region factorization. The approximation after the
``$\approx$'' corresponds to the $m^2/Q^2 \to 0$ limit of the ``$-$''
solution.

Expressing the plus and minus components in \eref{parlabels2} in terms
of rapidity, 
\begin{equation}
\inity = \frac{1}{2} \ln \parz{\left| \frac{Q^2}{\xnp^2 (\initq^2 + \initqt^2)} \right|} \, , 
\qquad \finaly = \frac{1}{2} \ln \parz{\left| \frac{\zexp^2 \Tscsq{q}{} + \delta \Tscsq{k}{} - 2 \zexp \T{q}{} \cdot \delta \T{k}{} + \finalq^2}{\zexp^2 Q^2} \right|} \, .
\end{equation}
Then, values of $\zexp$, $\xnp$, $\initq$, $\finalq$, $\initqt$,
$\finalqt$ can be mapped, along with values of $\ratgen$-$\ratX$, to
regions of a $\Tsc{q}{}$ versus rapidity map like \fref{Q2}.  If $\zexp \Tsc{q}{} = \order{Q}$, then $\finaly \approx \ln
\parz{\frac{\Tsc{q}{}}{Q}} \approx 0$, while if $\zexp \Tsc{q}{} =
\order{m}$, then $\finaly \approx \ln \parz{\frac{m}{Q}}$. 
In the handbag configuration, wherein all partonic
transverse momenta are zero, the parton four-momenta may be written,
\begin{equation}
\initq = \parz{\frac{\sqrt{-\initq^2}}{\sqrt{2}} e^{\inity}, -\frac{\sqrt{-\initq^2}}{\sqrt{2}} e^{-\inity}, \T{0}{} } \, , \qquad \finalq = \parz{\frac{\sqrt{\finalq^2}}{\sqrt{2}} e^{\finaly}, \frac{\sqrt{\finalq^2}}{\sqrt{2}} e^{-\finaly}, \T{0}{} } \, . 
\label{e.rapofpartons}
\end{equation}
Since $\initq^+ \approx -\qmom{+}{\rm b} = Q/\sqrt{2}$ and $\finalq^+
\approx \qmom{-}{\rm b} = Q/\sqrt{2}$ in the handbag configuration,
then $\inity \approx -\finaly = \order{\ln\parz{Q/m}}$.  Therefore,
partons in the handbag configuration are centered roughly on $y
\approx 0$ in the Breit frame.

Note also that if $\xn$ and $\zex$ are small, then according to
\eref{hadronraps} both the target and produced hadrons will tend to be
skewed toward larger rapidities in the Breit frame. Therefore, 
hadrons measured in the final state will tend to be at larger
rapidities than the corresponding handbag-configuration partons.

\section{Target Remnant Hadrons}
\label{s.target}

\begin{figure*}
\centering
\includegraphics[scale=0.6]{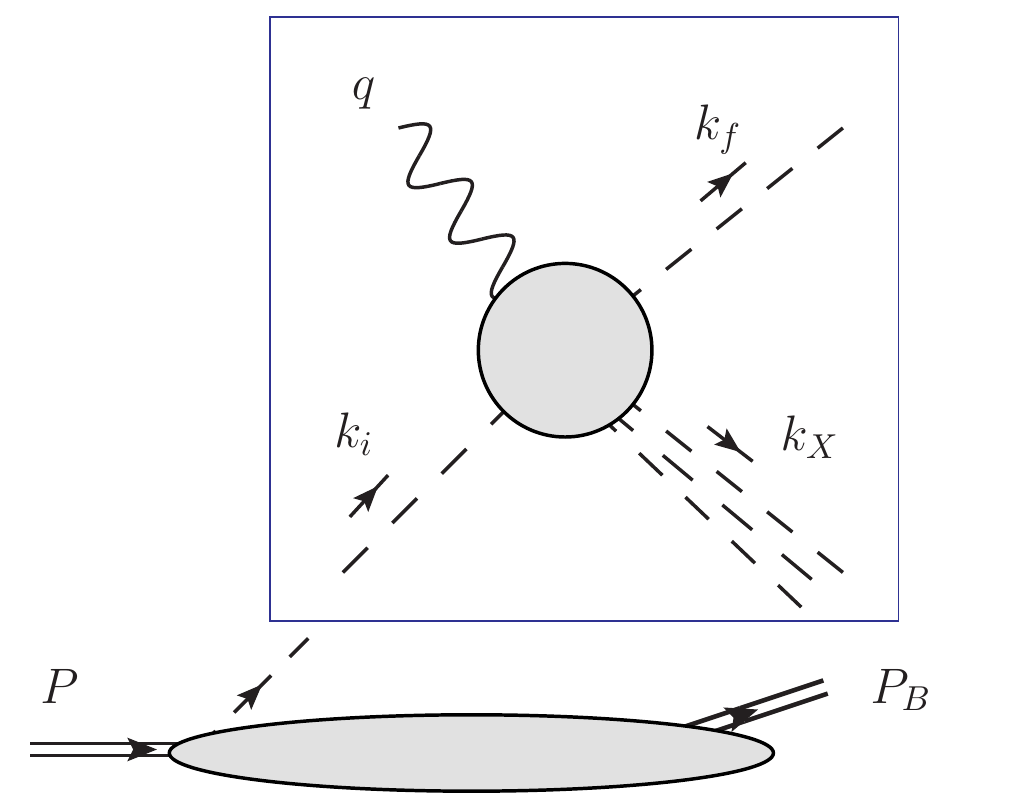}
\caption{A hadron produced in the target region -- see~\eref{tarreg2}. Hadrons produced from the hard part are not observed.}
\label{f.partonic_tar}
\end{figure*}

If, in contrast to the discussion in \sref{partons}, the hadron is in the target fragmentation region (see \fref{partonic_tar}), then
\begin{align}
\hadmomplain{}{} \cdot P \ll Q^2 \, , \label{e.tarreg2}
\end{align}
In the target region, $\zh$ is no longer as useful for parameterizing
the process since it no longer necessarily describes a momentum
fraction -- see \eref{zrels} and note that the quantity under the
square root diverges as $\zh \to 0$.  In terms of $\xh$, $\zex$ is:
\begin{align}
\zex 
&= \frac{\sqrt{4 \xbj^2 (\hadmass^2/Q^2)(1 - \Tscsq{q}{}/Q^2) + \xh^2} - \xh}{2 \xbj (1 - \Tscsq{q}{}/Q^2)} \nonumber \\
&= \frac{\hadmass^2 \xbj}{Q^2 \xh}-\frac{\hadmass^4 \xbj^3 \left(Q^2-\Tscsq{q}{} \right)}{Q^6 \xh^3} + \order{\frac{\hadmass^6 \left(Q^2-\Tscsq{q}{} \right)^2}{Q^{10}}} \, , 
\label{e.ztoxh}
\end{align}
where we have kept the solution that gives exactly $\zex = 0$ when
$\hadmomplain{}{}$ is exactly massless and collinear to $P$.  Now, 
\begin{equation}
\hadmomplain{}{} \cdot P = \frac{\pmass \hadtmass}{2} \parz{e^{\Delta y} + e^{-\Delta y}}
= \frac{\pmass^2 \xbj \left(\hadmass^2+\Tscsq{q}{} \zex^2 \right)}{Q \zex \left(\sqrt{4 \pmass^2 \xbj^2+Q^2}+Q\right)}+\frac{Q \zex \left(\sqrt{4 \pmass^2 \xbj^2+Q^2}+Q\right)}{4 \xbj} \, .\label{e.pdotphad}
\end{equation}
Equation~\eqref{e.pdotphad} is no larger than $\order{m^2}$ if $\zex
\sim m^2/Q^2$ and $\Tscsq{q}{} \zex^2/Q^2 \ll 1$.  So for the target
region, \eref{tarreg2} with \erefs{ztoxh}{pdotphad} means
\begin{equation}
\zex = \Theta \parz{\frac{m^2}{Q^2}} \, .
\end{equation}
The ``Big $\Theta$'' symbol is used because the first term in
\eref{pdotphad} puts a lower limit on acceptable sizes for $\zex$. In
other words, the target region criterion fails both when $\zex \ll
m^2/Q^2$ as well as when $\zex \gg m^2/Q^2$.  From \eref{ztoxh}, this
means the target fragmentation criterion in terms of $\xh$, $\xbj$ and
$\hadmomt{}{}$ is
\begin{equation}
\frac{\xh}{\xbj} = \order{1} \, , \qquad  \frac{\Tscsq{q}{} \zex^2}{Q^2} = \frac{\hadmomtsq{2}{b} }{Q^2}  \ll 1 \, .
\end{equation}
To translate \eref{tarreg2} into a dimensionless ratio, define
\begin{equation}
\rattar = \frac{\hadmomplain{}{} \cdot P}{Q^2} 
= \frac{\pmass^2 \xbj \left(\hadmass^2+\Tscsq{q}{} \zex^2 \right)}{Q^3 \zex \left(\sqrt{4 \pmass^2 \xbj^2+Q^2}+Q\right)}+\frac{\zex \left(\sqrt{4 \pmass^2 \xbj^2+Q^2}+Q\right)}{4 \xbj Q} \, . 
\label{e.R4}
\end{equation}
Therefore, the target region criterion is 
\begin{equation}
\rattar \ll 1 \, .
\end{equation}
In \cite{Boglione:2016bph}, it was $1/\ratcur$ that was used to
characterize the target region, and that is another acceptable
definition, but \eref{R4} has the advantage of working even when
$\initq$ differs significantly from $P$ and of being simpler to
calculate. 

\section{Soft-Central Hadrons}
\label{s.soft}

It is possible that for some hadrons, $\ratTM \ll 1$, while neither
$\ratcur$ nor $\rattar$ is small. We call this the soft region since
such hadrons are not a product of hard scattering but do not associate
in any obvious way with a quark or target direction. 

\section{Specific Examples}
\label{s.examples}
For illustration, let us insert some specific numbers into the above system of formulas. First, consider 
the purely kinematical ratios in \sref{kins} for realistic experimental scenarios. 
In \fref{Qx} we display the ($Q,\xbj$) kinematic coverage of three SIDIS experiments: JLab 12 (11~GeV 
electron beam), HERMES (27.5 GeV electron beam) and COMPASS (160 GeV muon beam).
The shaded regions are obtained by applying the appropriate experimental cuts in each case, as reported in Refs.~\cite{Avakian:2016rst,Airapetian:2012ki,Aghasyan:2017ctw}.
Notice that the JLab 12 kinematics covers a very wide range of $\xbj$ values, well above 0.6, but it is limited to intermediate/small values of $Q$. Instead, the COMPASS kinematics reaches up to much larger values of Q, but the accessible range of $\xbj$ is confined to values no larger than 0.4.   
In each plot, the values of the ratio $\xn/\xbj$, \eref{xndef}, are color coded: darker shades represent regions where $\xn/\xbj$ deviates from $1$ and the MTA approximation deteriorates. 
As expected mass corrections are more important at 
large values of $\xbj$ and small values of $Q$.

\fref{zPTzN} shows the ratio $\zex/\zh$, over the $(\zh,P_{B,T}/Q)$ kinematic coverage of the three experiments. Again darker shades represent larger deviations from $1$ which, in this case, are more significant than for $\xbj/\xn$, especially at JLab kinematics.

It is helpful to sketch the landscape of possible scenarios in a
transverse momentum versus rapidity map like the one shown
in~\fref{Q2}. Each of the regions discussed in \sref{partons}, \sref{target}, and \sref{soft} is 
represented there as a colored blob, and the task is to determine the sizes of the blobs, their borders, and their 
degree of overlap. The relevant power suppression factors are shown. (Recall, for example, \eref{R2}.)
\begin{figure}[htb]
\centering
\includegraphics[width=1.15\textwidth]{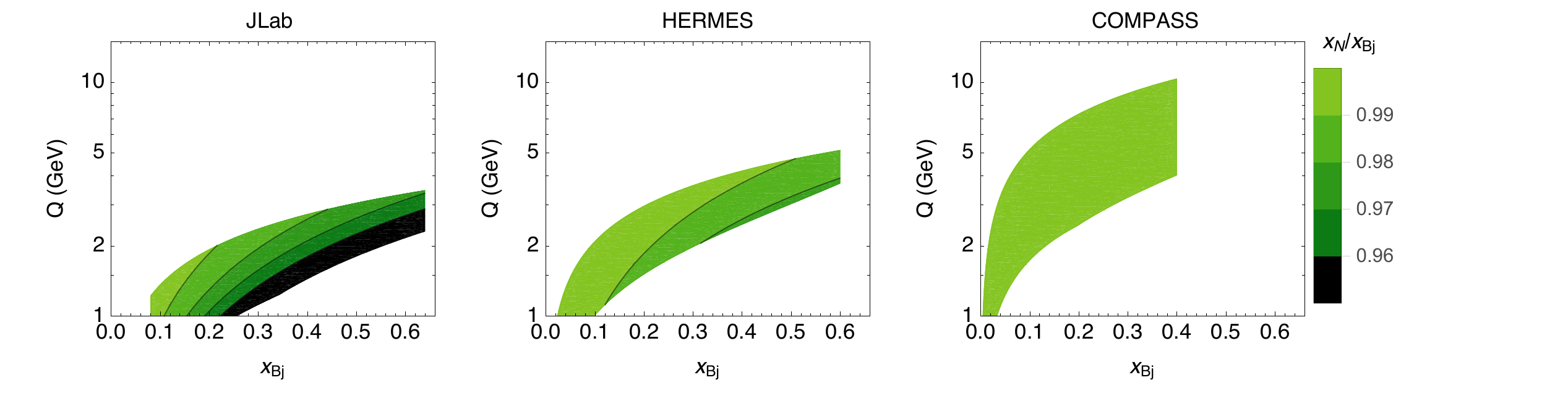}
\caption{The kinematic regions of $Q$ and $\xbj$ covered by JLab 12 (left
panel), HERMES (central panel) and COMPASS (right panel).
The shaded areas are obtained by applying the appropriate experimental
cuts in each case, as reported in
Refs.~\cite{Avakian:2016rst,Airapetian:2012ki,Aghasyan:2017ctw}.
These plots show that $Q$ and $\xbj$ are strongly correlated: large values
of $\xbj$ can only be accessed when $Q$ is sufficiently
large; conversely, when $Q$ is relatively small, only limited values of
$\xbj$ can be reached.
The values of $x_N/\xbj$, as obtained using~\eref{xndef}, are
color-coded: the lightest shade corresponds to values very close to one,
while darker shades correspond to regions where the ratio
$x_N/\xbj$ increasingly deviates from $1$ and the quality of the MTA
deteriorates.
Notice that, while mass corrections are more important for JLab 12 kinematics, in all cases considered $\xn/\xbj\approx1$ to good approximation. 
}
\label{f.Qx}
\end{figure}
\begin{figure}[htb]
\centering
\includegraphics[width=1.1\textwidth]{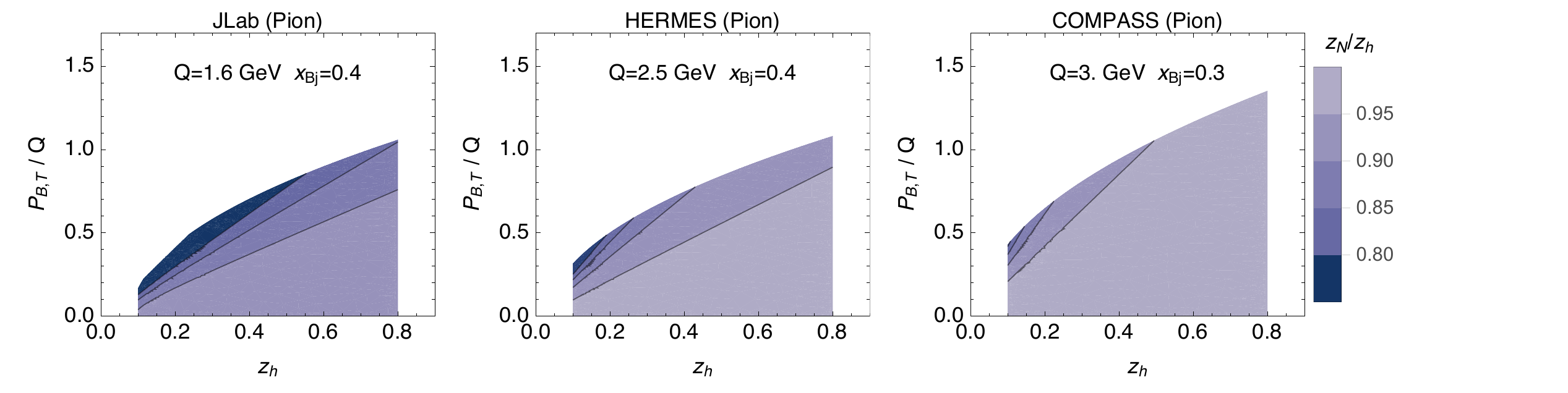}
\includegraphics[width=1.1\textwidth]{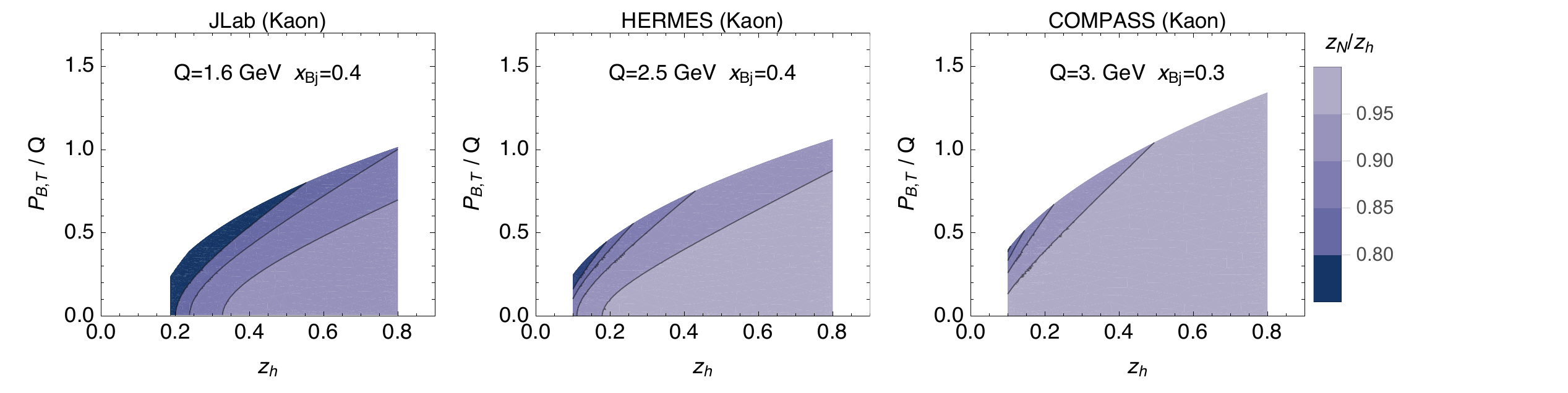}
\caption{The ratio $z_N/z_h$, \eref{zrels2}, is represented over the kinematic coverage in
($z_h,P_{B,T}/Q$)
for JLab 12 (left panels), HERMES (central panels) and COMPASS (right
panels), at some fixed values of $\xbj$ and $Q$,
as indicated in the plot title.
Appropriate experimental cuts, as reported in
Refs.~\cite{Avakian:2016rst,Airapetian:2012ki,Aghasyan:2017ctw}, are applied in each case.
The values of $z_N/z_h$, for pion production (upper panels) and kaon
production (lower panels)
are obtained using~\eref{zrels2} and are color-coded: the lightest shade
corresponds to values very close to one, while darker shades
correspond to regions where the ratio $z_N/z_h$
increasingly deviates from 1 and the quality of the MTA deteriorates. 
Notice how deviations from $1$ are more sizable
as compared to those of $\xn/\xbj$ in \fref{Qx}, particularly in the JLab case. 
}
\label{f.zPTzN}
\end{figure}
\begin{figure}[htb]
\centering
\includegraphics[width=0.9\textwidth]{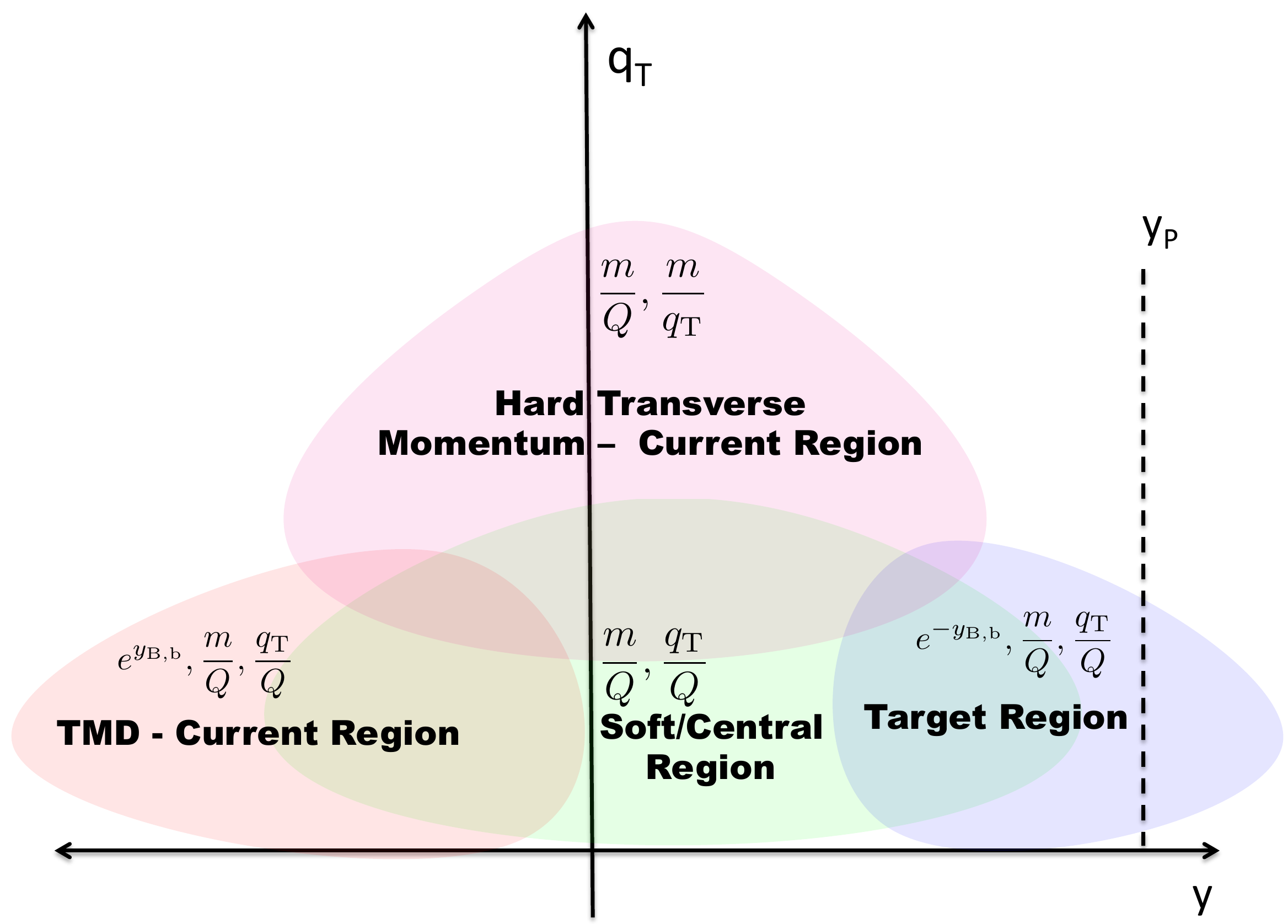}
\caption{
  Sketch, not to-scale, of kinematical regions of SIDIS in terms of
  the produced hadron's Breit frame rapidity and transverse momentum. In
  each region, the type of suppression factors that give factorization
  are shown.  (The exact size and shape of each region may be very
  different from what is shown and depends on quantities like $Q$ and
  the hadron masses.) In the Breit frame, according to
  \eref{rapofpartons}, partons in the handbag configuration are centered
  on $y \approx 0$ if $-\initq^2 \approx \finalq^2 = \order{m^2}$.  The shaded 
  regions in the sketch are shifted somewhat toward the target rapidity $\py$ (the
  vertical dashed line) to account for the behavior of \eref{hadronraps}
  when $\zex$ and $\xn$ are small.}
\label{f.Q2}
\end{figure}

To give more detailed examples than the above, a few assumptions about 
non-perturbative properties
of partons are necessary. $300$~MeV is a typical estimate of
non-perturbative mass scales so we try $\initq = \finalq = \delta
\Tsc{k}{} = 300$~MeV. Also, to start with we assume that $\T{q}{} \cdot
\delta \T{k}{} = \Tsc{q}{} \delta \Tsc{k}{}$. (Azimuthal
effects may be added later.)

In addition, the particular partonic kinematics of interest need to be
specified. Say, for example, that the goal is to examine target partons in the 
valence region (such as discussed on page 3 of \cite{Avakian:2016rst}).  Then the 
focus should be on momentum fraction values of $\xi$ roughly around $0.3$.  
For $\zeta$, we might reasonably focus on values where collinear fragmentation 
functions are large but have reasonably small uncertainties, say $\zeta \approx 0.3$. 
From Fig.~\ref{f.Qx}, JLab12 measurements at $\xbj \approx 0.2$ may 
reach to as large as about $2$~GeV in $Q$.

First let's consider overall kinematics. Contour plots
of $W_{\rm SIDIS}^2$, \eref{Wsidis}, are shown for a pion 
mass in \fref{kinplots} for (a) $\Tsc{q}{} = 0$ 
and (b) $\Tsc{q}{} = 2.0$~GeV, giving a sense of what is kinematically 
possible for the SIDIS remnant at different $\Tsc{q}{}$ and for 
lower $Q$. The expectation is that the 
area near the kinematically forbidden region, where the final state 
phase space vanishes, does not readily separate into distinct regions as in \fref{Q2}. So in the below we will focus on kinematics away from those boundaries. Also, for now we will restrict to large enough $Q$ that $\ratgen$ in \eref{R0} is
negligible, so $\ratcur$ is the first of the $\ratgen$-$\ratX$ that we will
consider here.
\begin{figure}[htb]
\centering
\includegraphics[width=1.1\textwidth]{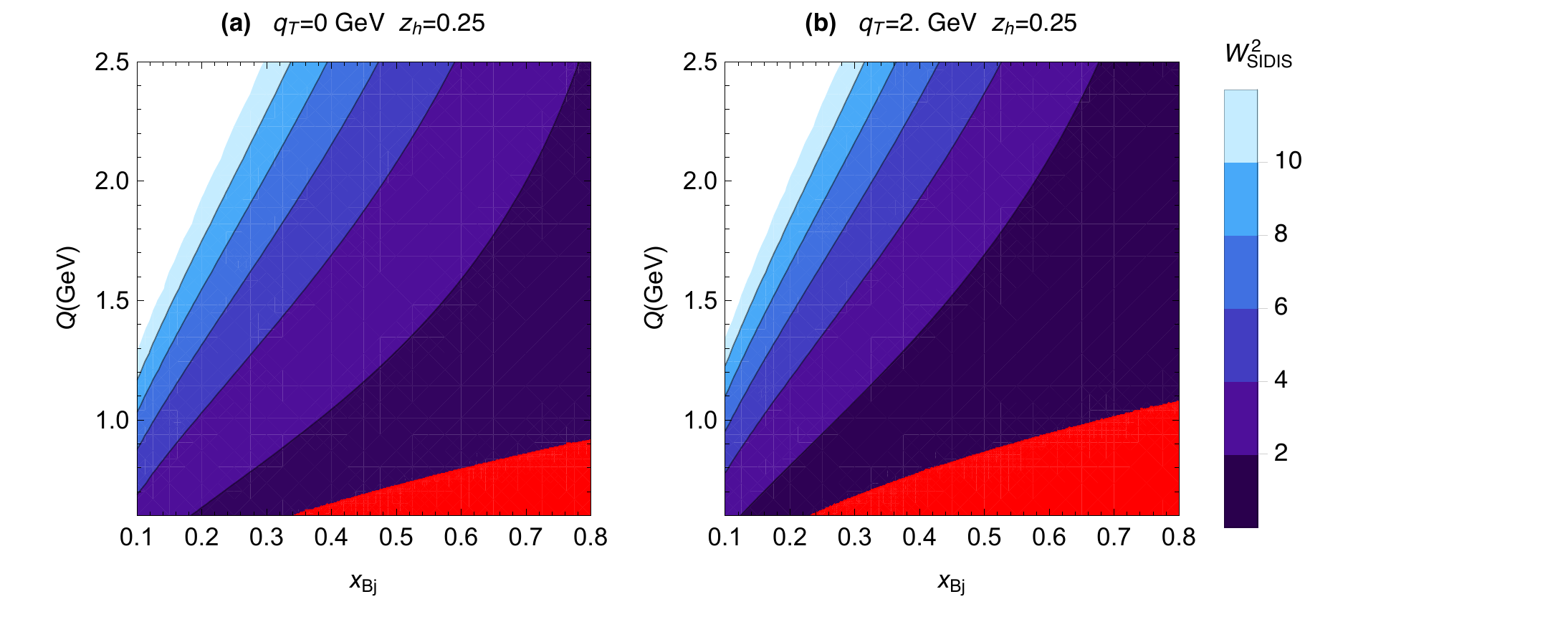}
\caption{
  Plots (a)-(b) show $W_{\rm SIDIS}^2$, \eref{Wsidis}, for $\Tsc{q}{} = 0$ and $\Tsc{q}{} = 2.0$~GeV 
  respectively for the case of a produced pion. $\zh = 0.25$ in each case. The red region is
  kinematically forbidden.  Near to the kinematically forbidden region, it is to be expected that 
  a clear separation into regions along the lines of \fref{Q2} will break down. The classification according 
  to the sizes of $\ratgen$-$\ratX$ is cleaner at larger $Q$ and with small but fixed $\xbj$. Note that the corresponding plots for a heavier final state hadron have a larger forbidden region.)
}
\label{f.kinplots}
\end{figure}

For the representative values discussed above ($\xi =
0.3$, $\zh = 0.25$, $\zeta = 0.3$ and a small $\Tsc{q}{} = 0.3$~GeV),
values of $\ratcur$ are shown on the $Q$ vs. $\xbj$ contour plot
in \fref{R1plots}. The trend is as expected: at large $Q$ and not-too-large $\xbj$, $\ratcur$ remains small for all transverse momenta, while corrections
might be necessary at smaller $Q$ and larger $\xbj$. 

In addition to
confirming the current-region approximation, which holds valid where collinearity $\ratcur$ is small, it is necessary to map out
the applicability of large and small transverse momentum
approximations. 
For this we turn to $\ratTM$. \fref{R2plots} is an example that corresponds
to the same kinematics as \fref{R1plots}. It confirms basic expectations, such as that what constitutes ``large-$\Tsc{q}{}$'' grows with $Q$. It also shows that, while the
hadron is in the current region for most $\Tsc{q}{}$ as in
\fref{R1plots} (a,b), the small
transverse momentum region shown in \fref{R2plots} (a) is much more restrictive.  For $\Tsc{q}{}
\lesssim 0.5$~GeV, $\ratTM$ is firmly in the small transverse momentum
region for most of the $Q$ shown, while for $\Tsc{q}{} \gtrsim 1.5$~GeV $\ratTM$ indicates that we are well in the large transverse momentum region. There is a
broad intermediate region where the situation is not clear.
The flavor of the final state hadron is a decisive factor in
determining the relevant factorization region. For example, comparing
the plots of $\ratcur$ in \fref{R1plots} for (a)  $M_B = m_\pi$, and (c) $M_B = m_K$, shows a completely different behaviour of the collinearity ratio $\ratcur$. For $ Q = 1.5$ GeV and $x_{Bj} =
0.1$, $\ratcur \approx 0.1$ for pions and $\ratcur \approx 0.8$ for
kaons. If $\ratcur \approx 0.8$ is taken to be large, then
confidence that one is in the current region deteriorates.
\begin{figure}[htb]
\includegraphics[width=1.1\textwidth]{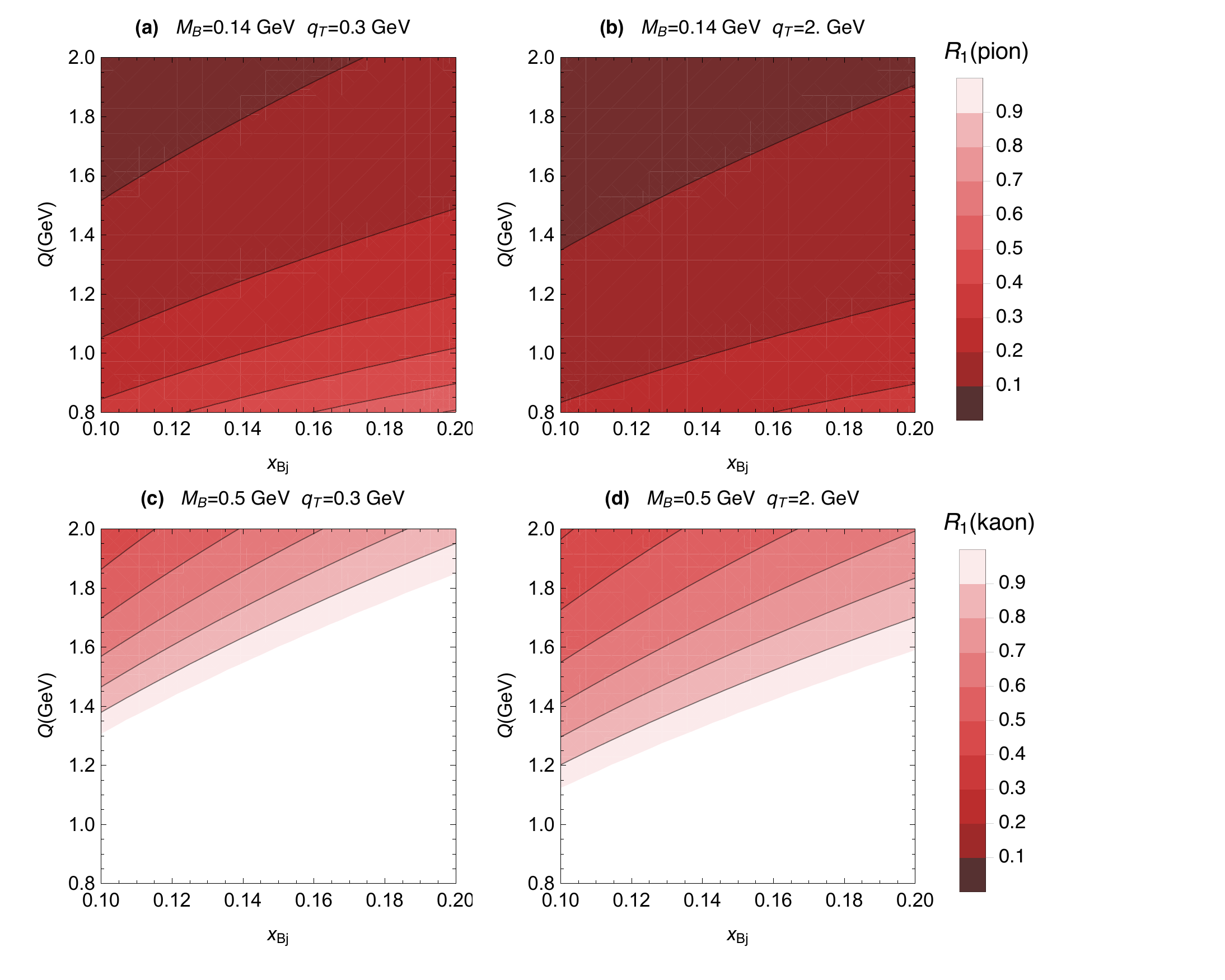}
\caption{
  Collinearity ($\ratcur$ from \eref{R1} for fixed $z_h=0.25$, $\zeta=0.3$ and $\xi=0.2$. Top panels show the ratio for
  $M_B = m_\pi$ at (a) small 
  transverse momentum ($\Tsc{q}{} = 0.3$~GeV) and (b) $\Tsc{q}{} = 2.0$~GeV. Similar cases for $M_B = m_K$ are shown in the bottom panels, (c) and (d). 
}
\label{f.R1plots}
\end{figure}
\begin{figure}[htb]
\centering
\includegraphics[width=1.\textwidth]{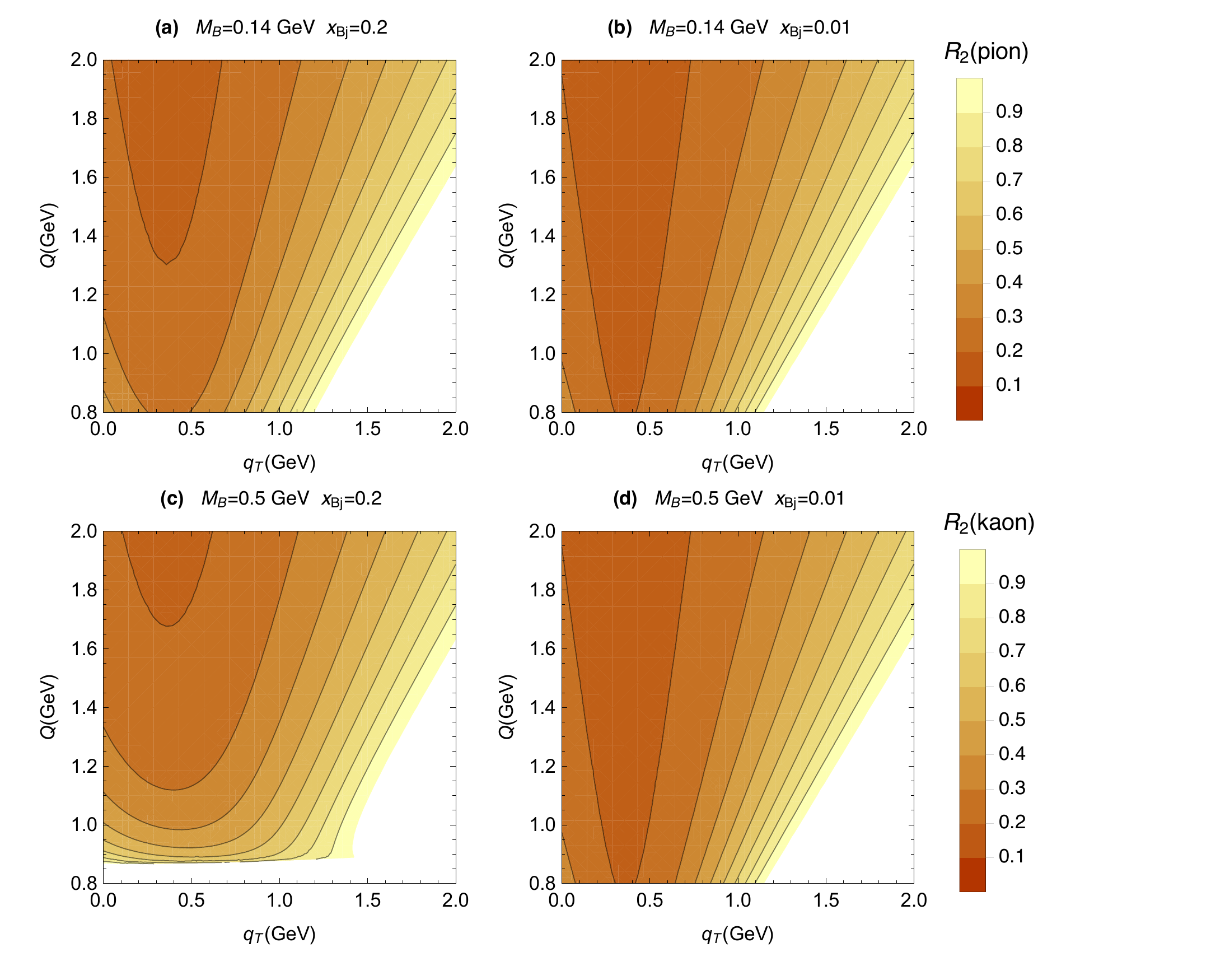}
\caption{
  Transverse momentum hardness, $\ratTM$, from \eref{ratTM} for fixed $z_h=0.25$, $\zeta=0.3$ and $\xi=0.2$. Top panels show the ratio for
  $M_B = m_\pi$ at (a) $\xbj=0.2$ and (b) $\xbj=0.01$. Similar cases for $M_B = m_K$ are shown in the bottom panels, (c) and (d). 
}
\label{f.R2plots}
\end{figure}
The flavor of the final state hadron has little effect on the transverse
momentum hardness, $\ratTM$, from \eref{ratTM}. From
\fref{R2plots} (a) and \fref{R2plots} (c) flavor dependence is only noticeable at low Q 
and even then the effect is small. To summarize, the produced hadron mass affects collinearity $\ratcur$
significantly, but does not appear to be a primary factor in determining transverse hardness $\ratTM$. 

Within a specific example, collinearity $\ratcur$ and transverse hardness $\ratTM$ have helped us to map out the current kinematic region (small $\ratcur$) and to separate the "small" from the "large" transverse momentum regions (small $\ratTM$ vs large $\ratTM$). The former will reasonably correspond to a region where we expect TMD factorization to apply, while for the latter a collinear factorization will be appropriate. At this stage, one might wonder whether a
LO calculation could be enough or whether higher order perturbative corrections are necessary. This is where $\ratX$ comes into the game: large $\ratX$ coupled with large $\ratTM$ signal a large $q_T$ region where presumably higher order pQCD corrections are  relevant, while small $\ratX$ together with small $\ratTM$ clearly indicate a TMD current region, which requires a TMD factorization scheme.

Clearly the above indications only apply to the specific example we have chosen, corresponding to specific values of the kinematic variables ($\xi =0.3$, $\zh = 0.25$, $\zeta = 0.3$, $\Tsc{q}{} = 0.3$~GeV) and of the non perturbative parameters ($\initq = \finalq = \delta\Tsc{k}{} = 300$~MeV, $\T{q}{} \cdot
\delta \T{k}{} = \Tsc{q}{} \delta \Tsc{k}{}$). A web tool which allows to compute $\ratcur$-$\ratX$ for any kinematic configuration can be found in Ref.~\cite{sidistool}.
%
\section{Conclusion}
\label{s.conclusion}

Since the early work in presented in Refs.~\cite{Meng:1995yn,Nadolsky:1999kb,Graudenz:1994dq}  there has been a large number of studies on unpolarized SIDIS cross sections~\cite{Koike:2006fn,Anselmino:2006rv,Anselmino:2013lza,Signori:2013mda,Sun:2013dya,Sun:2013hua,Bacchetta:2017gcc}. Unpolarized SIDIS is, however, only one component in a broad program of phenomenological studies where the universality of parton correlation functions plays a central role in testing pictures of nucleon structure~\cite{Anselmino:2016uie,Anselmino:2015sxa,Bacchetta:2015ora,Scimemi:2017etj,Kang:2017btw,Kang:2014zza,Landry:2002ix,Sun:2016kkh,Sun:2011iw,Echevarria:2015uaa,Guzzi:2013aja,Echevarria:2016scs,Echevarria:2014xaa,Boglione:2018dqd,Boglione:2017jlh,Kang:2015msa,Lin:2017stx,Ye:2016prn,Bacchetta:2019tcu,Bastami:2018xqd,Bertone:2019nxa}.   This demands a clear language for identifying kinematical regions of transversely differential deep inelastic scattering cross sections with particular underlying partonic pictures, especially in regions of moderate to low Q where sensitivity to kinematical effects outside the usual very high energy limit becomes non-trivial.

In this paper, we have outlined the ways that the questions about the boundaries between
different partonic regimes of SIDIS can be posed systematically, based
on the power-law expansions that apply in each region (recall
\fref{Q2}). As the ratios $\ratgen$-$\ratX$ in \sref{partons} show, quantifying the separation 
between different SIDIS regions requires at least some rough model assumptions for 
the intrinsic properties of partons. Hence, our position is that region mapping  
should be viewed as one of the aspects of SIDIS that is to be determined with 
guidance from data, rather than being treated as well-known input. 
Nevertheless, the $\ratgen$-$\ratX$ can already be useful 
for querying the reasonableness of some region assumptions. For example, if collinearity $\ratcur$ is 
found to be approximately $10$ for a wide range of even rough models, then a current region 
assumption could be viewed with skepticism.
Conversely, very small values of collinearity $\ratcur$  might be considered a strong signal that one is deep in a regime 
where a current region fragmentation function picture is appropriate. If, in addition, there is a small transverse hardness ratio $\ratTM$ it may be taken to signal  the close proximity to small transverse momentum, where a TMD factorization scheme would be appropriate. If transverse hardness ratio $\ratTM$ and spectator virtuality ratio $\ratX$ are both large, then high order pQCD corrections are likely important.
In a fitting context, the 
$\ratgen$-$\ratX$ can be utilized to fix Bayesian priors. Conversely, the success or 
failure of theoretical predictions can be used to constrain the ranges of 
$\ratgen$-$\ratX$ that are acceptable for particular regions in 
future theoretical predictions.  

In developing a picture of the likelihood that a
particular kinematical region corresponds to a particular partonic
picture, one should of course consider a wide range of multiple non-perturbative
models for the values of $\initq$, $\finalq$, etc., in addition to
sampling from a range of $\zeta$, $\xi$, and azimuthal angles, and
track the values of $\ratgen$-$\ratX$, in addition to $\xn/\xbj$,
$\zex/\zh$, $W_{\rm tot}^2$, $W_{\rm SIDIS}^2$ to assess the validity
of various purely kinematical approximations.  This could be done,
perhaps, at the level of computer simulations, where the values of
$\ratgen$-$\ratX$ can be tracked. For now, the effect of changing quantities like $\initq^2$ and 
$\finalq^2$ can be examined directly with our web tool Ref.~\cite{sidistool}.

In the future we plan to incorporate this view into phenomenological procedures, 
particularly in situations with not-too-large $Q$. We hope that this will ultimately 
contribute to a clearer picture of the borders between different regions and an 
improved understanding of the transition between hadronic and partonic degrees of 
freedom.

\begin{acknowledgments}	
We thank Eric~Moffat, Wally~Melnitchouk and Jian-Wei Qiu for useful discussions. We are especially thankful to Markus Diehl for discussions of \eref{esumrule} that led to \aref{bdgmms}.
T.~Rogers, A.~Dotson, S.~Gordon and N.~Sato were supported by the U.S. Department of Energy, Office of 
Science, Office of Nuclear Physics, under Award Number DE-SC0018106. This work 
was also supported by the DOE Contract No. DE- AC05-06OR23177, under which 
Jefferson Science Associates, LLC operates Jefferson Lab. 
L. Gamberg was supported by the U.S. Department of Energy under grant No. DE-FG02-07ER41460. 
A.~Prokudin was supported by the National Science Foundation under Grant No.~PHY-1623454 and the DOE Contract No. DE- AC05-06OR23177, under which Jefferson 
Science Associates, LLC operates Jefferson Lab.
J.~O.~Gonzalez-Hernandez work was partially supported
by Jefferson Science Associates, LLC under  U.S. DOE Contract DE-AC05-06OR23177 and by the 
U.S. DOE Grant DE-FG02-97ER41028.
\end{acknowledgments}

\appendix 

\section{Integration over $\zex$ and $\T{P}{B,}{}$.}
\label{a.bdgmms}
In the appendix, we will work in the target hadron rest frame (a photon frame). Start 
with the elementary relation
\begin{equation}
\sum_B 
\int \diff{^2 \hadmomt{\gamma}} \, \diff{\zex}{}  
\left( \frac{\diff{\sigma^B}{}}{\diff{\xbj}{} \diff{y}{} 
\diff{\psi}{} \diff{^2 \hadmomt{\gamma}} \, \diff{\zex}{}}  \right) 
= \langle N  \rangle \frac{\diff{\sigma^{\rm tot}}{}}{\diff{\xbj}{} \diff{y}{} \diff{\psi}{} } \, .
\end{equation}
Change the $\zex$ variable on the left side to $\zh$. The $\diff{\zh}{}$ appears in 
both the integral and the derivative and Jacobian factors cancel:
\begin{equation}
\sum_B 
\int \diff{^2 \hadmomt{\gamma}} \, \diff{\zh}{}  
\left( \frac{\diff{\sigma^B}{}}{\diff{\xbj}{} \diff{y}{} 
\diff{\psi}{} \diff{^2 \hadmomt{\gamma}} \, \diff{\zh}{}}  \right) 
= \langle N  \rangle \frac{\diff{\sigma^{\rm tot}}{}}{\diff{\xbj}{} \diff{y}{} \diff{\psi}{} } \, .
\end{equation}
Expressed in differential form, and for one particular hadron type $B$, this is
\begin{equation}
 \diff{^2 \hadmomt{\gamma}} \, \diff{\zh}{}  
 \left( \frac{\diff{\sigma^B}{}}{\diff{\xbj}{} \diff{y}{} \diff{\psi}{} 
 \diff{^2 \hadmomt{\gamma}} \, \diff{\zh}{}}  \right) 
 = \diff{\langle N_B  \rangle} 
 \frac{\diff{\sigma^{\rm tot}}{}}{\diff{\xbj}{} \diff{y}{} \diff{\psi}{} } \, , \label{e.differ}
\end{equation}
where $\diff{\langle N_B  \rangle}$ is the number of particles of type $B$ in the 
differential volume $ \diff{^2 \hadmomt{\gamma}} \, \diff{\zh}{}$.  Let $E_B$ be the 
energy \emph{per particle} of type $B$ (in the target rest frame), and multiply both 
sides of \eref{differ} by 
$E_B$:
\begin{align}
E_B  \diff{^2 \hadmomt{\gamma}} \, 
\diff{\zh}{}  
\left( \frac{\diff{\sigma^B}{}}{\diff{\xbj}{} \diff{y}{} \diff{\psi}{} 
\diff{^2 \hadmomt{\gamma}} \, \diff{\zh}{}}  \right) 
& = E_B \diff{\langle N_B  \rangle} 
\frac{\diff{\sigma^{\rm tot}}{}}{\diff{\xbj}{} \diff{y}{} \diff{\psi}{} } \nonumber \\
& = \diff{\langle E_B^{\rm all}  \rangle} 
\frac{\diff{\sigma^{\rm tot}}{}}{\diff{\xbj}{} \diff{y}{} \diff{\psi}{} }\, . \label{e.differ2}
\end{align}
$E_B \diff{\langle N_B  \rangle}$ is the energy \emph{per} $B$-particle \emph{times} 
the number of $B$ particle in the differential volume, so it is the total energy of 
\emph{all} $B$-particles in the differential volume. Therefore, we have defined it as 
$\diff{\langle E_B^{\rm all} \rangle}$ in the last equality. Integrating it and summing 
over all types of final state particles produces the total energy of the entire final 
state:
\begin{equation}
\sum_B \int \diff{\langle E_B^{\rm all}  \rangle} = E^{\rm tot} \, . \label{e.energysum}
\end{equation}
Note that the sum over $B$ is a sum over all \emph{types} of particles, not a sum 
over actual particles. Divide both sides of \eref{differ2} by $q^0$:
\begin{equation}
\frac{E_B}{q^0}  \diff{^2 \hadmomt{\gamma}} \, 
\diff{\zh}{}  
\left( \frac{\diff{\sigma^B}{}}{\diff{\xbj}{} \diff{y}{} \diff{\psi}{} 
\diff{^2 \hadmomt{\gamma}} \, \diff{\zh}{}}  \right) 
= \frac{1}{q^0} \diff{\langle E_B^{\rm all}  \rangle} 
\frac{\diff{\sigma^{\rm tot}}{}}{\diff{\xbj}{} \diff{y}{} \diff{\psi}{} }\, . \label{e.integranddiffer3}
\end{equation}
Integrate over both sides, restore the sum over particle types $B$, and use 
\eref{energysum} for the right side:
\begin{equation}
\sum_B \int \frac{E_B}{q^0}  \diff{^2 \hadmomt{\gamma}} \, 
\diff{\zh}{}  
\left( \frac{\diff{\sigma^B}{}}{\diff{\xbj}{} \diff{y}{} \diff{\psi}{} 
\diff{^2 \hadmomt{\gamma}} \, \diff{\zh}{}}  \right) 
= \frac{E^{\rm tot} }{q^0} 
\frac{\diff{\sigma^{\rm tot}}{}}{\diff{\xbj}{} \diff{y}{} \diff{\psi}{} }\, . \label{e.differ3}
\end{equation}
Now, in the target rest frame,
\begin{equation}
\zh = \frac{P \cdot \hadpsc}{P \cdot q} = \frac{E_B}{q^0} \, .
\end{equation}
Also, 
\begin{equation}
q^0 = \frac{Q^2}{2 \pmass \xbj}, \qquad P^0 = \pmass \, .
\end{equation}
From energy conservation, 
\begin{equation}
E^{\rm tot} = q^0 + P^0 \, ,
\end{equation}
so 
\begin{equation}
\frac{E^{\rm tot} }{q^0} = \frac{q^0 + P^0 }{q^0} 
= 1 + 2 \xbj \pmass^2/Q^2 = \parz{1 + \frac{\gamma^2}{2 \xbj}}\, .
\end{equation}
So, \eref{differ3} becomes
\begin{equation}
\sum_B \int \zh  \diff{^2 \hadmomt{\gamma}} \, 
\diff{\zh}{}  \left( \frac{\diff{\sigma^B}{}}{\diff{\xbj}{} \diff{y}{} \diff{\psi}{} 
\diff{^2 \hadmomt{\gamma}} \, \diff{\zh}{}}  \right) 
=  \parz{1 + \frac{\gamma^2}{2 \xbj}} \frac{\diff{\sigma^{\rm tot}}{}}{\diff{\xbj}{} 
\diff{y}{} \diff{\psi}{} }\, . \label{e.differ4}
\end{equation}

Now we need to use this to relate the SIDIS and the total DIS structure functions. 
For the total DIS cross section, the structure function decomposition with standard 
notational conventions uses 
\eref{sidis1}, \eref{hadronictensor}, and \eref{incstructdec}. The cross section is thus
\begin{align}
\label{e.unpolstruct2b}
\frac{\diff{\sigma^{\rm tot}}{} }{\diff{\xbj}{} \diff{y}{} \diff{\psi}{}} & = 
\frac{\alpha_{\rm em}^2 y}{\xbj Q^2 (1 - \kinep)} \bigl[  F^{\rm tot}_T 
+ \kinep F^{\rm tot}_L \bigr] \, . 
\end{align}
Substituting \eref{unpolstruct2b} into the right side of \eref{differ4}, and substituting  
\eref{unpolstruct4} into the left side gives
\begin{equation}
\int \diff{\zh}{} \diff{^2 {\hadmomt{b}{}} }{} \zh \frac{1}{4 \zex} 
\frac{\xn \left(\sqrt{1
-\frac{4 \pmass^2 \xbj^2 M_{\rm B, T}^2}{Q^4 \zh^2}}+1\right)}{2 \xbj \sqrt{1
-\frac{4 \pmass^2 \xbj^2 M_{\rm B, T}^2}{Q^4 \zh^2}}} F_{T/L} 
= \left( 1 + \frac{\gamma^2}{2 \xbj} \right) F^{\rm tot}_{T/L} \, .
\end{equation}
Substituting \eref{zrels2} for $\zex$ gives the factor in \eref{unpolstruct4}.
Thus, the normalization of $F_{T/L}$ needs to be redefined as in \eref{barred} 
in order to get the integration/sum rule in \eref{esumrule} 
and \cite[Eqs.~(2.18-2.21)]{Bacchetta:2006tn}.

\bibliography{bibliography}

\end{document}